\documentclass[
reprint,
amsmath,amssymb,
aps,
]{revtex4-2}

\usepackage{graphicx}
\usepackage{xcolor}
\usepackage{dcolumn}
\usepackage{bm}
\usepackage{siunitx}
\usepackage{hyperref}

\hypersetup{
    colorlinks=true,
    linkcolor=blue,
    citecolor=blue,
    filecolor=blue,
    urlcolor=blue
}

\newcolumntype{d}[1]{D{.}{.}{#1}}
\newcommand\T{\rule{0pt}{2.6ex}}       
\newcommand\B{\rule[-1.2ex]{0pt}{0pt}} 

\graphicspath{{./}{./figs/}}

\begin{document}

\preprint{APS/123-QED}

\title{Accurate, precise pressure sensing with tethered optomechanics}

\author{O. R. Green}
\altaffiliation[Present address: ]{Physics Department, Case Western Reserve University, Cleveland, OH 44106}
\author{Y. Bao}
\author{J. R. Lawall}
\author{J. J. Gorman}
\author{D. S. Barker}
\email{daniel.barker@nist.gov}
\affiliation{National Institute of Standards and Technology, Gaithersburg, MD 20899}

\date{\today}

\begin{abstract}

We show that optomechanical pressure sensors with characterized density and thickness can achieve uncertainty as low as \(1.1~\si{\percent}\) via comparison with a secondary pressure standard.
The agreement between the secondary standard and our optomechanical sensors is a necessary step towards using optomechanical devices as primary pressure sensors.
Our silicon nitride and silicon carbide sensors are short-term and long-term stable, displaying Allan deviations compatible with better than \(1~\si{\percent}\) precision and baseline drift significantly lower than the secondary standard.
Our measurements also yield the \textit{in situ} thin-film density of our sensors with \(1~\si{\percent}\) total uncertainty or lower, aiding development of other optomechanical sensors.
Our results demonstrate that optomechanical pressure sensors can achieve accuracy, precision, and drift sufficient to replace high performance legacy pressure gauges.

\end{abstract}

\maketitle


\section{\label{sec:intro}Introduction}


Sensors based on mechanical damping and deflection have a long history in precision pressure and vacuum metrology~\cite{Fremerey1985, hyland1985}.
Recently, the principles of such mechanical sensors have been applied in optomechanical systems.
The size and sensitivity of optomechanics allows new measurement paradigms, such as extreme squeezed-film enhancement~\cite{salimi2024} or direct detection of individual gas molecule collisions~\cite{Magrini2021,barker2024}.
When the mass and area of the device are known, optomechanical devices offer pressure measurements that do not rely on calibration by a reference pressure gauge~\cite{Scherschligt2018}.
Such measurements have been demonstrated with levitated~\cite{Blakemore2020} and tethered~\cite{reinhardt2024} devices at approximately \(10~\si{\percent}\) uncertainty.
However, the demonstrated accuracy and precision of optomechanical pressure sensors to date are insufficient to replace high performance gauges (with typical accuracy of \(1~\si{\percent}\) or better) in industrial and metrological applications.

Here, we investigate the limits to the accuracy, precision, and long-term stability of tethered silicon nitride and silicon carbide optomechanical pressure sensors.
We focus on whether optomechanical pressure sensors can produce accurate readings without calibration by a pressure standard. 
In pressure metrology, calibration is the process of comparing the readings of a pressure sensor to a pressure standard and subsequently correcting the readings to ensure that the sensor is accurate to within its uncertainty~\cite{JCGMVIM3}.
A pressure standard is a device with known accuracy and quantified uncertainty that measures or generates a pressure.
There are two main types of pressure standard: primary and secondary.
A primary pressure standard derives its accuracy and uncertainty from physical models and characterization measurements that are independent of pressure (\textit{i.e.}, it is accurate without being calibrated by another pressure standard).
A secondary pressure standard derives its accuracy from a calibration by a primary pressure standard.
Therefore, we can determine how accurate an optomechanical pressure sensor is without calibration by comparing its readings to a pressure standard and observing that those readings are accurate to within their uncertainty without correction.

\begin{figure*}
    \centering
    \includegraphics[width=\textwidth]{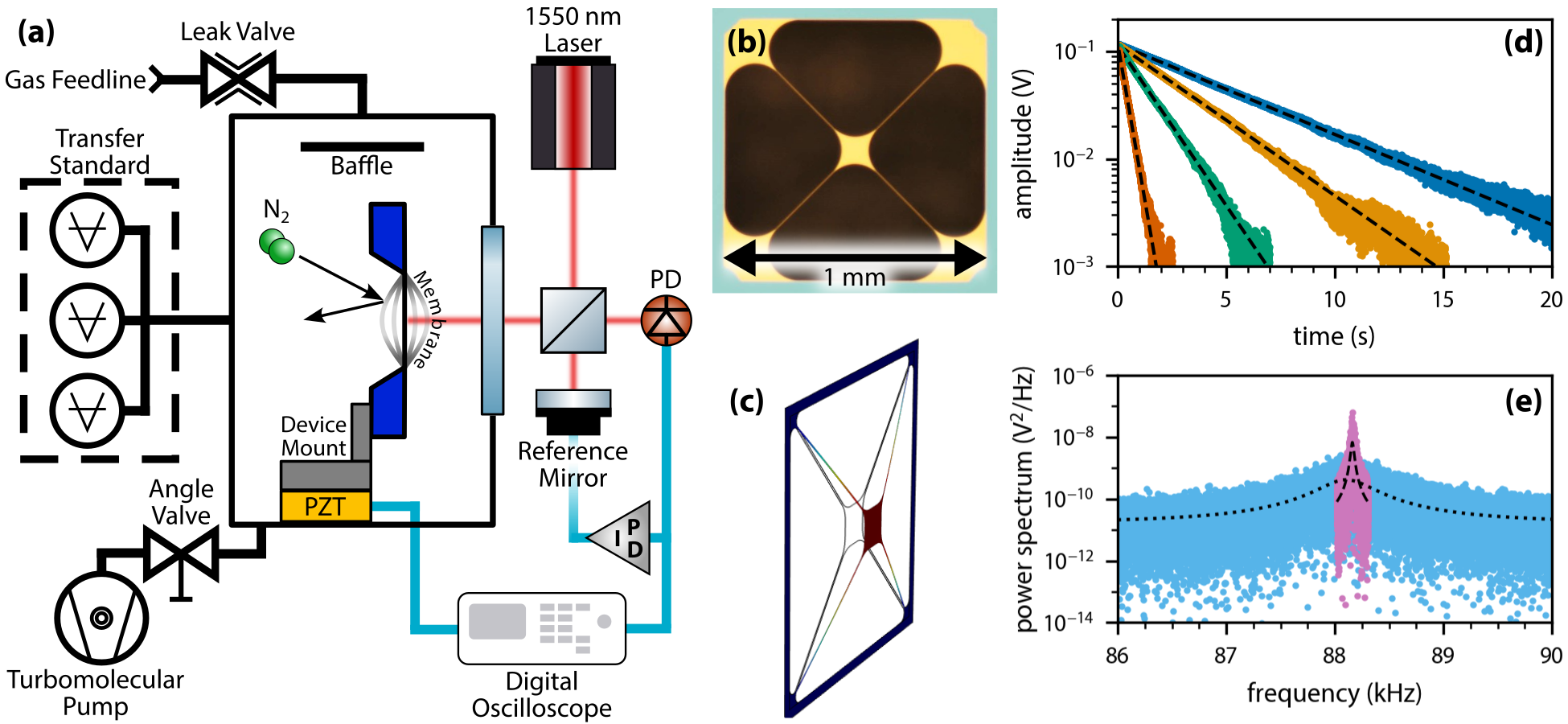}
    \caption{(a) Schematic of the experimental apparatus.
    (b) Optical microscope image of a representative trampoline similar to SiN~1.
    (c) Simulation of the fundamental mode of an optomechanical trampoline.
    (d) Typical mechanical ring-down signals for SiN~1.
    Single, Hilbert-transformed ring downs are shown at base pressure (blue), \(10~\si{\milli\pascal}\) of Ar (orange), \(50~\si{\milli\pascal}\) of Ar (green), and \(260~\si{\milli\pascal}\) of Ar (red).
    Black dashed curves are exponential fits to each ring-down.
    (e) Typical thermo-mechanical noise spectra for SiN~1.
    Single spectra are shown at \(11~\si{\pascal}\) of Ar (pink), and \(291~\si{\pascal}\) of Ar (light blue).
    Black dashed and dotted curves are Lorentzian fits to the lower pressure spectrum and higher pressure spectrum, respectively.}
    \label{fig:apparatus}
\end{figure*}

The pressure standard in our study is directly calibrated by the National Institute of Standards and Technology's (NIST) primary pressure standards (\textit{i.e.}, it is a secondary pressure standard).
Our optomechanical sensors use their gas-collision-induced mechanical damping response to measure the pressure via pressure-independent characterization of their density and thickness.
By comparing the pressure reported by our optomechanical sensors and the secondary standard, we demonstrate that optomechanical pressure measurements can achieve accuracies competitive with high precision legacy gauges in an operating range from \(10^{-2}~\si{\pascal}\) to \(10~\si{\pascal}\)~\cite{Comsa1980, Fremerey1985, jousten2021}.
The operating range overlaps with the high vacuum range (\(10^{-6}~\si{\pascal}\) to \(0.1~\si{\pascal}\)) and medium vacuum range (\(0.1~\si{\pascal}\) to \(100~\si{\pascal}\)), which are important in semiconductor fabrication.
The high vacuum range is of particular interest in vacuum metrology because recently developed cold-atom standards and refractometry standards do not currently operate over the entire high vacuum range~\cite{egan2015, egan2016, barker2023}.

Our results show that our optomechanical devices are consistent with NIST's primary pressure standards and that our devices are accurate without correcting their readings via calibration when measuring heavy gases (relative molecular mass \(M_r\ge 28\)).
This is an important step toward using optomechanical devices as primary pressure sensors.
By analogy with the definition of a ``primary standard'' above, we define a primary pressure sensor as a sensor that is accurate to within a quantified uncertainty without calibration by a pressure standard.

The agreement between optomechanical devices and NIST's standards is not restricted to a particular sensor material or geometry.
Our measurements show that  silicon nitride trampolines and silicon carbide membranes are both well-described by mechanical damping models derived from the kinetic theory of gases~\cite{Cavalleri2010, Martinetz2018}.
We find that our single-crystal silicon carbide devices reach total pressure measurement uncertainties of approximately \(1~\si{\percent}\) while our amorphous silicon nitride sensors are limited to approximately \(5~\si{\percent}\) uncertainty.

Our apparatus and data acquisition techniques are described in Sec.~\ref{sec:apparatus}.
We detail the operating principle of optomechanical pressure sensors based on mechanical damping and our methods for comparing the optomechanically measured pressure to the secondary standard in Sec.~\ref{sec:thry}.
Section~\ref{sec:res} presents the results of our study and we conclude in Sec.~\ref{sec:conclusion}.
Appendix~\ref{sec:devices} describes fabrication of our silicon nitride devices and Appendix~\ref{sec:characterization} details our pressure-independent characterization of all our devices.
We derive the mechanical damping response of our sensors in Appendix~\ref{sec:eq1_der} and provide details on our fitting routines in Appendix~\ref{sec:fitting}.
Appendix~\ref{sec:uncertainty} contains uncertainty analysis for our optomechanical sensors and secondary standard.

\section{\label{sec:apparatus}Apparatus}

We assess the pressure-sensing performance of optomechanical devices by placing them in a vacuum chamber whose pressure is continuously measured by a secondary standard.
The main features of the apparatus are shown in Fig.~\ref{fig:apparatus}(a).
We have sequentially installed three optomechanical devices in the apparatus.
By comparing devices made from different materials with different resonator geometries to the secondary standard, we can test the consistency of the mechanical damping model (see Sec.~\ref{sec:thry}) over variation in device design.
Devices SiN~1 and SiN~3 are silicon nitride trampolines fabricated at NIST~\cite{Reinhardt2016, Norte2016}.
Device SiN~1 has a \(1~\si{\milli\meter}\) wide, square support frame and a photonic crystal mirror patterned into its central pad.
Device SiN~3 has a \(3~\si{\milli\meter}\) wide, square support frame.
It does not have a photonic crystal etched in its central pad.
A representative silicon nitride trampoline is shown in Fig.~\ref{fig:apparatus}(b).
The fundamental mechanical modes of SiN~1 and SiN~3, see Fig.~\ref{fig:apparatus}(c), have resonance frequencies of approximately \(88~\si{\kilo\hertz}\) and approximately \(30~\si{\kilo\hertz}\), respectively.
Further details of the silicon nitride device fabrication are provided in Appendix~\ref{sec:devices}.
Device SiC~2 is a commercial silicon carbide membrane manufactured by Norcada, Inc~\cite{disclaimer}.
Device SiC~2 is a \(2~\si{\milli\meter}\) wide, square membrane. 
The resonance frequency of SiC~2's \((1,3)\) mode is approximately \(217~\si{\kilo\hertz}\).
The devices are attached to an in-vacuum piezoelectric shaker.

As we shall see in Sec.~\ref{sec:thry}, the sensitivity of each device to pressure variation is determined by its density \(\rho\) and thickness \(h\).
The effect of \(\rho\) and \(h\) on the device pressure sensitivity is captured by the pressure sensitivity coefficient \(S=1/\rho h\).
We predict \(S\) for each device from pressure-independent measurements of \(\rho\) and \(h\), which we describe in Appendix~\ref{sec:characterization}.
Table~\ref{tab:devpars} shows \(h\), \(\rho\), and \(S\) with the associated standard uncertainty (at coverage factor \(k=1\)) for each device.
(The coverage factor \(k\ge 1\) multiplies the uncertainty \(u_X\) in a quantity \(X\) to establish a level of confidence that the true value of \(X\) lies within \(\pm k u_X\) of the measured value of \(X\).
When the underlying distributions are approximately Gaussian, \(k=1\) corresponds to a \(68~\%\) level of confidence and \(k=2\) corresponds to a \(95~\%\) level of confidence~\cite{JCGMGUM}.)

\begin{table}
    \caption{Summary of relevant characteristics for each device.
    The thickness \(h\), density \(\rho\), and sensitivity \(S=1/\rho h\) are measured as described in Appendix~\ref{sec:characterization}.
    Parenthetical quantities represent the standard uncertainty at coverage factor \(k=1\) (see text).}
    \label{tab:devpars}
    \begin{ruledtabular}
        \begin{tabular}{cccc}
            Device\T & \(h~(\si{\nano\meter})\) & \(\rho~(\si[per-mode=symbol]{\kilo\gram\per\cubic\meter})\) & \(S~(\si[per-mode=symbol]{\square\meter\per\kilo\gram})\)\B \\
            \hline
            SiN~1\T & \(220.4(0.6)\) & \(2810(141)\) & \(1615(81)\) \\
            SiN~3\T & \(224.2(1.3)\) & \(2810(141)\) & \(1587(80)\) \\
            SiC~2\T & \(51.3(0.5)\) & \(3210(20)\) & \(6074(66)\) \\
        \end{tabular}
    \end{ruledtabular}
\end{table}

We measure the motion of our optomechanical devices with a homodyne Michelson interferometer.
The photodiode signal measuring the interferometer fringe is sent to a digital oscilloscope and to a proportional-integral (PI) controller.
The digital oscilloscope records intensity fluctuations at the mechanical resonance frequency.
The PI controller stabilizes the Michelson interferometer fringe via low-frequency feedback to the position of a reference mirror.

The vacuum system is evacuated to a base pressure of approximately \(2\times 10^{-5}~\si{\pascal}\) by a \(77~\si{\liter\per\second}\) turbomolecular pump.
We introduce ultra-high purity test gases into the vacuum chamber with a variable leak valve.
A baffle located in front of the gas inlet ensures that the test gas thermalizes with the vacuum chamber walls before interacting with any of our pressure sensors.
Two calibrated industrial platinum resistance thermometers (PRTs) measure the vacuum system temperature.
The PRTs are attached to opposite sides of the vacuum chamber exterior, with one near the gas inlet and the other at the connection to the secondary standard.

The secondary standard is a suite of calibrated capacitance diaphragm gauges (CDGs).
The CDGs are directly calibrated by NIST's oil and mercury ultrasonic interferometer manometers.
We discuss the measurement uncertainties of the CDGs in Appendix~\ref{sec:unc_cdg}.
A calibrated PRT monitors the secondary standard temperature and reports that it is stable to better than \(\pm0.01~\si{\celsius}\) over the duration of our measurement campaign.
Further description of the secondary standard is available in Ref.~\cite{miller1997,ricker2017}.

We measure the damping rate of our devices using mechanical ring-down and the thermo-mechanical noise spectrum.
For mechanical ring-down, we resonantly excite the device under test using the piezoelectric shaker.
The excited mechanical motion ``rings down'' when the resonant drive is removed, yielding an exponentially decaying oscillation of the Michelson interferometer signal.
We Hilbert transform the signal to extract the envelope, which we fit to an exponential decay to measure the total mechanical damping rate \(\Gamma_t\).
Figure~\ref{fig:apparatus}(d) shows typical Hilbert-transformed ring-downs with the associated exponential fits.

To measure thermo-mechanical noise spectrum, we digitize the interferometer signal without applying an excitation to measure the Brownian motion of the device.
We then construct the power spectrum of the mechanical motion from the signal's fast Fourier transform.
A Lorentzian fit to the power spectrum at the mechanical resonance frequency then yields \(\Gamma_t\).
Figure~\ref{fig:apparatus}(e) shows example spectra with Lorentzian fits.

The mechanical ring-down and thermo-mechanical noise measurements agree within their statistical uncertainty.
At low pressure, mechanical ring-down offers higher signal-to-noise ratio than thermo-mechanical noise.
At high pressures, the increased gas damping degrades the achievable ring-down excitation amplitude.
We therefore switch from using ring-down to thermo-mechanical noise above a transition pressure of \(5~\si{\pascal}\).

\section{\label{sec:thry}Theory of Operation}

Mechanical oscillators are subject to pressure-dependent damping due to collisions with ambient gas.
The damping rate exhibits three regimes as a function of pressure.
At sufficiently low pressures, the oscillator is in the ``intrinsic damping regime'', where the gas-induced damping is small compared to at least one other damping mechanism.
The intrinsic damping in our devices is likely dominated by bending losses and anchor losses~\cite{Chakram2014, Norte2016, Tsaturyan2017}.
In the intrinsic damping regime, the total mechanical damping rate is roughly equal to the intrinsic damping rate \(\Gamma_0\), which includes all pressure-independent damping processes.
At higher pressures, the oscillator enters the ``molecular-flow-dominated regime'', where gas-induced damping is stronger than intrinsic damping and the total mechanical damping rate depends linearly on pressure~\cite{Christian1966, newell1968, Cavalleri2010, Martinetz2018}.
At yet higher pressures, the oscillator reaches the ``viscous-flow-dominated regime'', where the mean free path of the gas molecules becomes comparable to, or smaller than, the size of the oscillator{\color{red} ,} and the total mechanical damping rate increases with the square root of the pressure~\cite{hosaka1995, lubbe2011}~\footnote{Eventually, the size of the chamber enclosing the oscillator will also affect the damping rate, see Ref.~\cite{reich1982, Fremerey1985}.}.

The gas-induced mechanical damping rate is caused by fluctuations in the force from the gas on the mechanical oscillator due to collisions.
Because the particle-surface scattering physics is complicated, the collisions are described by a phenomenological model in which gas particles that collide with the mechanical oscillator scatter either specularly or diffusely from the oscillator surface~\cite{RamsayMolBeams, Cavalleri2010, Martinetz2018}.
In specular scattering, the gas particle collides elastically with the oscillator surface and its kinetic energy is conserved.
The gas particle therefore always transfers twice its initial normal momentum to the mechanical oscillator~\cite{Christian1966, Cavalleri2010, Martinetz2018}.
In diffuse scattering, the gas particle thermalizes with the oscillator surface before being emitted according to a cosine distribution.
In general, a gas particle that diffusely scatters does not transfer twice its initial normal momentum to the mechanical oscillator, which leads to a reduced mean force fluctuation compared to specular collisions~\cite{Cavalleri2010, Martinetz2018}.
The momentum accommodation coefficient \(\alpha\in [0,1]\) parametrizes the ratio of diffuse to specular collisions, and is defined as the fraction of molecules that reflects diffusely from the oscillator surface.
The remaining fraction of the gas (\(1-\alpha\)) reflects specularly.

In molecular flow, the gas-induced damping rate can be calculated from the force noise power spectral density of the gas particle impacts using the fluctuation-dissipation theorem~\cite{Cavalleri2010, Martinetz2018}.
Specular and diffuse collisions lead to different gas-induced mechanical damping rates due to their different mean force fluctuations.
For a thin mechanical oscillator, combining the specular and diffuse damping rates into a general expression gives~\cite{Cavalleri2010, Martinetz2018}
\begin{equation}
    \label{eq:mol_flow}
    \Gamma_m = \frac{(1+\pi/4)\alpha+2(1-\alpha)}{\rho h}\sqrt{\frac{8 m_g}{\pi k_B T}} P,
\end{equation}
where \(\rho\) is the oscillator density, \(h\) is the oscillator thickness, \(m_g\) is the molecular mass of the gas, \(P\) is the gas pressure, \(T\) is the temperature of the gas and oscillator (assumed to be in thermal equilibrium), and \(k_B\) is the Boltzmann constant.

Equation~\eqref{eq:mol_flow} depends on the mechanical oscillator geometry only through the thickness \(h\).
The independence of \(\Gamma_m\) from the mechanical oscillator's transverse size and shape is exact when the oscillator has only two surfaces -- a top and a bottom -- that gas molecules can collide with, which is the case for membranes like SiC~2.
The trampolines SiN~1 and SiN~3 have sidewalls exposed to the gas, which cause additional pressure-dependent damping.
We present a derivation of Eq.~\eqref{eq:mol_flow} that is particularly appropriate to membranes or trampolines in Appendix~\ref{sec:eq1_der}.
Our derivation shows that small corrections to Eq.~\eqref{eq:mol_flow} due to oscillator sidewalls exposed to the gas are negligible at our level of uncertainty.

By combining \(\Gamma_0\), \(\Gamma_m\), and the viscous damping rate \(\Gamma_v\), we arrive at the total mechanical damping rate~\cite{lubbe2011}
\begin{equation}
    \label{eq:tot_damp}
    \Gamma_t = \Gamma_0 + \frac{\Gamma_m \Gamma_v}{\Gamma_m+\Gamma_v}.
\end{equation}
When \(\Gamma_0 << \Gamma_m << \Gamma_v\), Eq.~\eqref{eq:tot_damp} simplifies to \(\Gamma_t \approx \Gamma_m\). 
A measurement of \(\Gamma_t\) thus yields the gas pressure through inversion of Eq.~\eqref{eq:mol_flow}. 

The analytical damping models of Equation~\eqref{eq:mol_flow} and Eq.~\eqref{eq:tot_damp} allow us to determine the requirements for optomechanical pressure sensing without calibration by a pressure standard.
We proceed by deriving the measurement equation for the pressure in the molecular flow limit, \(\Gamma_m << \Gamma_v\).
We find
\begin{equation}
    \label{eq:meas_p}
    P = \frac{\rho h}{(1+\pi/4)\alpha+2(1-\alpha)}\sqrt{\frac{\pi k_B T}{8 m_g}}\Gamma'_t,
\end{equation}
where \(\Gamma'_t = \Gamma_t-\Gamma_0\) is the background-subtracted, total mechanical damping rate.
Because \(\Gamma'_t\) depends on \(\Gamma_0\), the optomechanically measured pressure \(P\) is the pressure rise above background.
Sensors based on Eq.~\eqref{eq:meas_p} are therefore particularly suitable for process control applications.
To operate an optomechanical pressure sensor without calibration, all quantities aside from \(\Gamma'_t\) on the right-hand side of Eq.~\eqref{eq:meas_p} must be determined independently from \(P\) with quantified uncertainty~\cite{JCGMVIM3}.
Our apparatus and device characterization yield pressure independent values of \(\rho\), \(h\), \(T\), and \(m_g\) (see Sec.~\ref{sec:apparatus}, Appendix~\ref{sec:characterization}, and Appendix~\ref{sec:uncertainty}). 


\begin{figure*}
    \centering
    \includegraphics[width=\textwidth]{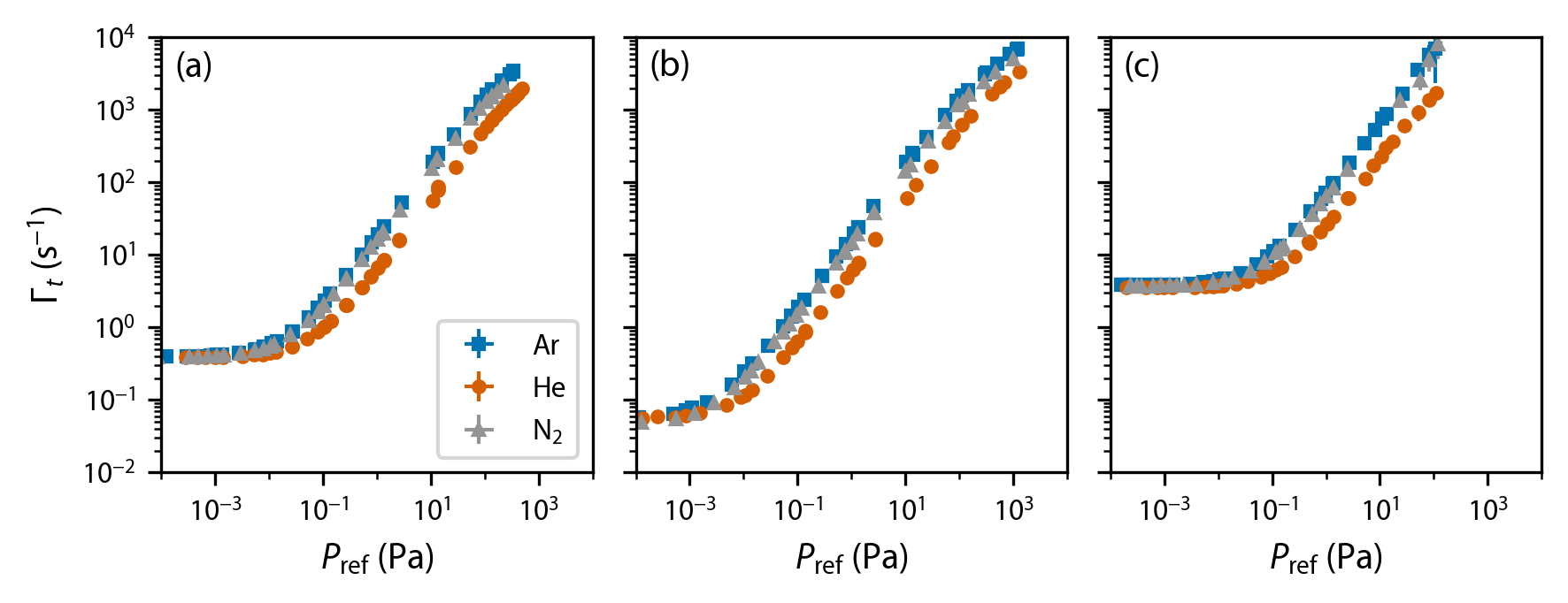}
    \caption{\(\Gamma_t\) as a function of \(P_{\rm ref}\) for device SiN~1 (a), SiN~3 (b), and SiC~2 (c).
    Blue squares, red circles, and gray triangles show data for Ar, He, and N\(_2\) test gases, respectively.
    Horizontal and vertical error bars denote the statistical standard uncertainty at coverage factor \(k=1\) (most error bars are smaller than the data points).
    }
    \label{fig:gamvp}
\end{figure*}

The only quantity that remains to be determined in Eq.~\ref{eq:meas_p} is \(\alpha\).
Gases principally reflect diffusely from surfaces, since the thermal de Broglie wavelength is typically small compared to the surface roughness~\cite{RamsayMolBeams, Comsa1980, Fremerey1985}.
Thus, \(\alpha=1\) is the only value that can be taken without requiring calibration by a pressure standard.
The pressure reading of a sensor based on linear damping (Eq.~\eqref{eq:meas_p}) varies by less than \(12~\si{\percent}\) between the limiting cases \(\alpha = 0\) and \(\alpha=1\).
This variation constrains the contribution to the uncertainty arising from \(\alpha\) to less than \(12~\si{\percent}\).
Prior studies of optomechanical gauges that damp according to Eq.~\ref{eq:mol_flow} have assumed that \(\alpha=0\)~\cite{Christian1966, lubbe2011, Scherschligt2018, reinhardt2024}, which limits their fractional accuracy to \(12~\si{\percent}\) unless they have been calibrated.

We test the uncertainty and operating pressure range where an optomechanical sensor provides accurate readings without calibration by a pressure standard.
We compute the optomechanical sensor's pressure reading via Eq.~\eqref{eq:meas_p} with \(\alpha=1\) and compare with the reading \(P_{\rm ref} \) of a secondary pressure standard.
We determine the pressure range in which the optomechanically inferred pressure
\begin{equation}
    \label{eq:pvp}
    P = \frac{S^{-1}}{1+\pi/4}\sqrt{\frac{\pi k_B T}{8 m_g}}\Gamma'_t 
\end{equation}
is equal to the background-subtracted secondary standard pressure
\begin{equation}
    P'_{\rm ref} = P_{\rm ref}-P_{\text{ref,}\,0},
\end{equation}
where \(P_{\text{ref,}\,0}\) is the secondary standard reading at base pressure (when no test gas has been introduced).
Because the secondary standard calibration is traceable to a primary pressure standard, the optomechanical pressure sensor is consistent with the primary pressure standard for the pressure range in which the readings agree within their mutual uncertainty. 

\section{\label{sec:res}Results}

We measure the damping response of our optomechanical sensors from approximately \(10^{-4}~\si{\pascal}\) up to approximately \(1~\si{\kilo\pascal}\).
Figure~\ref{fig:gamvp} shows a log-log plot of \(\Gamma_t\) for each sensor as a function of \(P_{\rm ref}\) for three test gases: Ar, He, and N\(_2\).
At low pressure, \(\Gamma_0\) makes the largest contribution to \(\Gamma_t\).
At intermediate pressure, the molecular flow damping exceeds \(\Gamma_0\) and the sensors exhibit a linear damping response to pressure.
At high pressure, the gas enters viscous flow and the slope diminishes as \(\Gamma_t\) transitions toward a square root dependence on pressure.
The onset of viscous flow is more apparent in Fig.~\ref{fig:gamvp}(a) and Fig.~\ref{fig:gamvp}(b) than in Fig.~\ref{fig:gamvp}(c), because the highest tested pressure is lower for the silicon carbide sensor shown in Fig.~\ref{fig:gamvp}(c).
The higher sensitivity of the silicon carbide sensor and the signal-to-noise ratio of the interferometer limit how far into the viscous flow regime we can measure the response of SiC~2.

\begin{figure*}
    \centering
    \includegraphics[width=\textwidth]{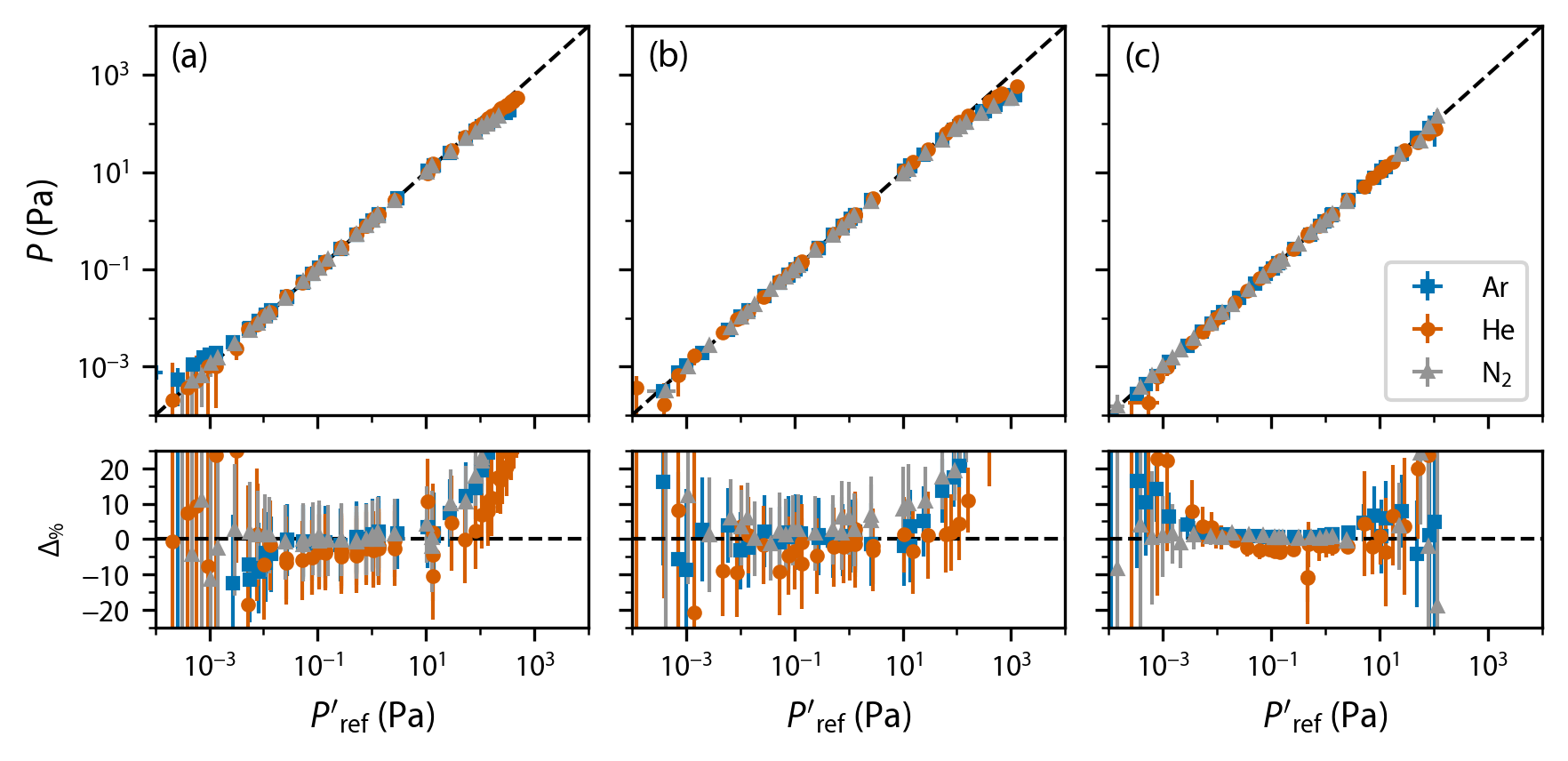}
    \caption{Pressure sensing performance of device SiN~1 (a), SiN~3 (b), and SiC~2 (c).
    In each subplot, the upper row shows \(P\) as a function of \(P'_{\rm ref}\) and the lower row shows the percent difference \(\Delta_{\%}\) between the two pressures.
    Blue squares, red circles, and gray triangles show data for Ar, He, and N\(_2\) test gases, respectively.
    Black dashed lines are a guide to the eye indicating perfect agreement between the optomechanical sensor and secondary standard.
    Horizontal and vertical error bars denote the total (statistical and non-statistical) standard uncertainty at \(k=1\).
    In the lower row, the vertical error bars incorporate the uncertainty in both \(P\) and \(P'_{\rm ref}\).
    In the upper row, most error bars are smaller than the data points.
    }
    \label{fig:pvp}
\end{figure*}

\subsection{\label{sec:prim_op}Pressure Measurement}

We assess each device's performance as a pressure sensor by comparing its pressure reading to the secondary standard.
We background subtract both the damping rate and the secondary standard pressure to eliminate contributions from residual gases and intrinsic damping.
We then convert \(\Gamma'_t\) to \(P\) using Eq.~\ref{eq:pvp}.
The optomechanical pressure \(P\) and secondary standard pressure \(P'_{\rm ref}\) are plotted against each other in the upper row of Figure~\ref{fig:pvp}.
In molecular flow, the optomechanical sensor and secondary standard readings collapse toward a line with unit slope.
The error bars in the upper row of Fig.~\ref{fig:pvp} represent the total (statistical and non-statistical) standard uncertainty (\(k=1\)) in both the optomechanical sensor and the secondary standard (see Appendix~\ref{sec:uncertainty} for a full uncertainty analysis).
The agreement between the sensor and standard is easier to discern through the percent difference \(\Delta_{\%} = 100\times(1-P/P'_{\rm ref})\), which is shown for each optomechanical sensor in the lower row of Fig.~\ref{fig:pvp}.
The uncertainty in \(\Delta_{\%}\) is propagated from both \(P\) and \(P'_{\rm ref}\).

For the heavy gases Ar and N\(_2\), all optomechanical sensors agree with the secondary standard within the mutual uncertainty in the molecular flow regime.
The agreement between optomechanical sensors and secondary standard occurs under the assumption that \(\alpha=1\), indicating that optomechanical devices are consistent with NIST's primary pressure standards.
We also find that the agreement between our sensors and the secondary standard is maintained over variation in mechanical oscillator geometry, material, and resonance frequency.
This result confirms that sidewall damping corrections to the silicon nitride trampoline response are negligible at our level of mutual uncertainty (see Appendix~\ref{sec:eq1_der}). 

The useful operating range for the optomechanical sensors extends roughly from \(10^{-2}~\si{\pascal}\) to \(10~\si{\pascal}\).
The lower limit of the operating range is the pressure at which the statistical uncertainty in \(\Gamma'_t\) becomes comparable to the uncertainty in \(S\).
The upper limit of the operating range is set by the transition into viscous flow.
Within the operating range, the total uncertainty of the optomechanical sensor is dominated by statistical uncertainty in \(\Gamma'_t\) and uncertainty in the device sensitivity \(S=1/\rho h\).
Combining these two uncertainties in quadrature, we find that device \{SiN~1, SiN~3, SiC~2\} is accurate to \{\(5.3~\si{\percent}\), \(6.1~\si{\percent}\), \(1.1~\si{\percent}\)\} for ring-down measurements in its operating range.
The accuracy of SiC~2 is competitive with high performance gauges operating in this pressure range and suggests that silicon nitride devices could achieve similar accuracy with better pressure-independent characterization of \(\rho\) (\textit{e.g.}, Rutherford backscattering~\cite{markwitz1993, huszank2016}).
For SiN~1 and SiN~3, the density uncertainty \(u_\rho\) dominates the total measurement uncertainty estimated above (\(u_\rho/\rho = 5~\si{\percent}\)).

The optomechanical sensors report a higher pressure for He than for Ar or N\(_2\).
Across all our sensors, the increase in reported He pressure is roughly \(3~\si{\percent}\) in the useful operating range. 
For SiN~1 and SiN~3, the He pressure still agrees with the secondary standard within the mutual uncertainty (see Fig.~\ref{fig:pvp} lower row).
However, the He pressure measured by SiC~2 disagrees with the secondary standard by more than twice the total standard uncertainty.

One plausible explanation for the shift in the He measurements is \(\alpha\ne 1\).
Specular reflections could contribute significantly to the damping due to helium's large thermal de~Broglie wavelength of approximately \(51~\si{\pico\meter}\) at \(295~\si{\kelvin}\)~\cite{RamsayMolBeams}.
Argon and N\(_2\) have smaller thermal de Broglie wavelengths of approximately \(16~\si{\pico\meter}\) and \(19~\si{\pico\meter}\), respectively.
A momentum accommodation coefficient for He scattering of \(\alpha_{\rm He}\approx 0.75\) would explain the shift from the measurements with heavier gases.
A lower momentum accommodation for He compared to Ar and N\(_2\) is also consistent with the thermal accommodation measurements of Refs.~\cite{trott2011,sharipov2016}.
Our observation of imperfect momentum accommodation for He gas suggests that optomechanical pressure sensors cannot take \(\alpha=1\) for light gases and will therefore be limited to approximately \(12~\si{\percent}\) uncertainty (see Sec.~\ref{sec:thry}).
Lower uncertainty pressure measurements of light gases may require calibration by a pressure standard.

\begin{figure}
    \centering
    \includegraphics[width=\linewidth]{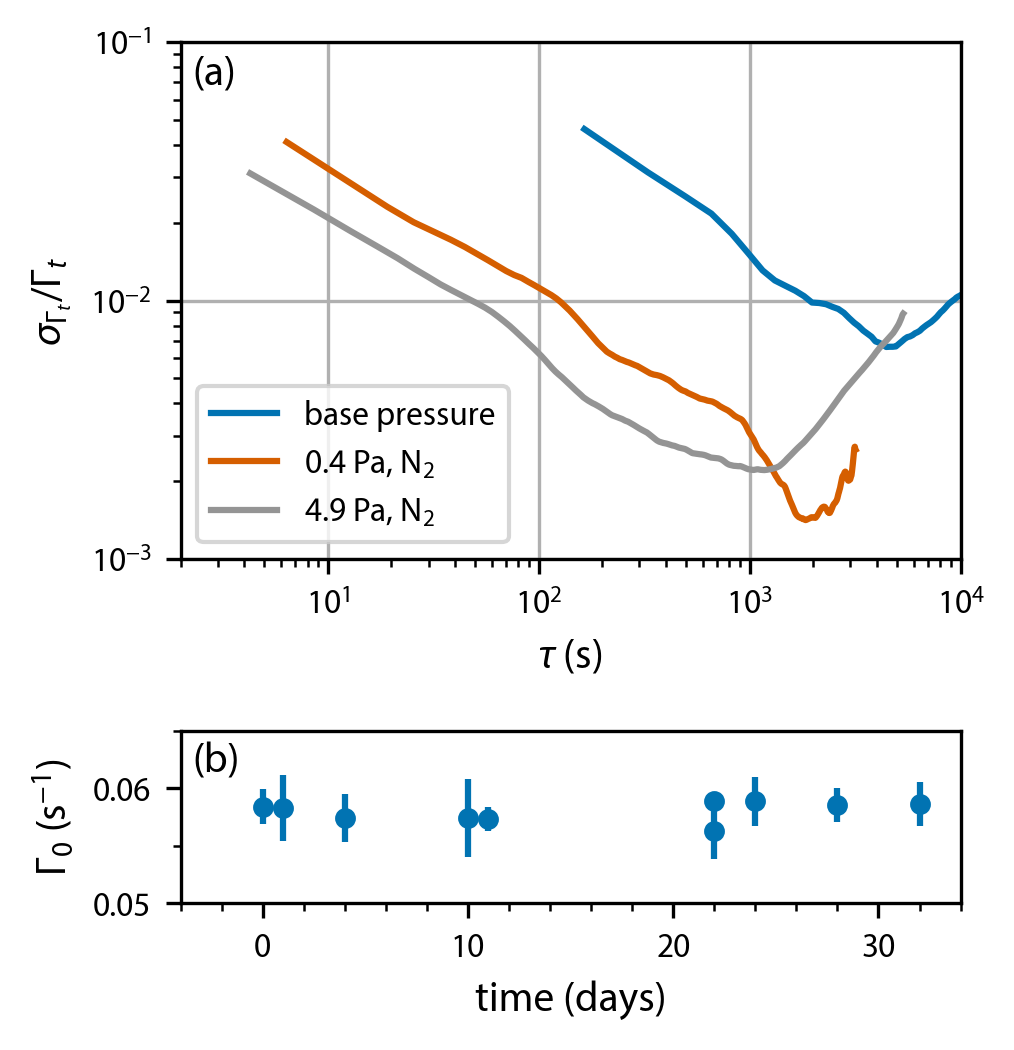}
    \caption{(a) Fractional Allan deviation \(\sigma_{\Gamma_t}/\Gamma_t\) as a function of averaging time \(\tau\) for SiN~3.
    The blue curve shows data taken at base pressure, while the red and gray curves show data for two pressures of \(N_2\) test gas.
    (b) Intrinsic damping \(\Gamma_0\) as a function of measurement day.
    Error bars represent the statistical uncertainty of \(5\) repeated measurements.
    }
    \label{fig:allan}
\end{figure}

\subsection{\label{sec:stability}Stability}

The utility of any gauge depends critically on its stability and baseline drift.
We characterize the stability using the fractional Allan deviation \(\sigma_{\Gamma_t}/\Gamma_t\).
We show \(\sigma_{\Gamma_t}/\Gamma_t\) as a function of averaging time \(\tau\) in Figure~\ref{fig:allan}(a) for ring-down measurements on device SiN~3 at three nominal pressures.
The minimum of \(\sigma_{\Gamma_t}/\Gamma_t\) is the lowest statistical uncertainty achievable with SiN~3 near the nominal pressure.
At base pressure, \(\sigma_{\Gamma_t}/\Gamma_t\) reaches a minimum of approximately \(0.7~\si{\percent}\).
The minimum occurs at an averaging time of approximately \(5000~\si{\second}\), which sets the maximum useful averaging time for SiN~3 in the current apparatus.
When we inject N\(_2\) test gas, \(\sigma_{\Gamma_t}/\Gamma_t\) reaches a minimum of approximately \(0.2~\si{\percent}\) after approximately \(1000~\si{\second}\) of averaging.
We suspect that the shorter useful averaging time for the N\(_2\) measurements is due to fluctuations of the leak valve conductance, which is supported by the behavior of the Allan deviation of the secondary standard (not shown).  

The Allan deviation measurements show that SiN~3 is stable for substantially longer than the averaging times used in Sec~\ref{sec:prim_op}.
The minimum \(\sigma_{\Gamma_t}/\Gamma_t\) at base pressure corresponds to a resolvable pressure rise of approximately \(40~\si{\micro\pascal}\) of N\(_2\) test gas at \(20~\si{\celsius}\).
The useful operating range of SiN~3 could therefore be extended down to approximately \(100~\si{\micro\pascal}\) at the expense of measurement time.
The high stability of SiN~3 also further supports our supposition that, with a lower uncertainty measurement of \(\rho\), silicon nitride devices could achieve total uncertainty comparable to those made from silicon carbide.


The baseline drift of our optomechanical sensors is lower than the pressure drift of the secondary standard.
We evaluate the baseline drift in our optomechanical sensors by repeatedly measuring their damping rate at base pressure under controlled laboratory conditions. 
Figure~\ref{fig:allan}(b) shows \(\Gamma_0\) for SiN~3 as a function of the time in days since the beginning of our measurement campaign (the baseline drift of SiN~1 and SiC~2 were evaluated over shorter time intervals, see Appendix~\ref{sec:unc_optomech}).

On a day-to-day basis, all successive \(\Gamma_0\) measurements agree within their mutual uncertainty.
The drift per day between successive measurements has 
an unweighted average magnitude of \(\overline{|\Gamma_0|}=2.5(1.5)\times 10^{-3}~\si{\per\second}/\)d.
(Here, and throughout the paper, parenthetical quantities represent \(k=1\) standard uncertainties~\cite{JCGMGUM}.)
The drift in \(\Gamma_0\) is consistent with zero to within two standard uncertainties. 
The drift magnitude \(\overline{|\Gamma_0|}\) is equivalent to an N\(_2\) pressure drift of approximately \(160~\si{\micro\pascal}/\)d at \(20~\si{\celsius}\), which is more than \(10\) times lower than the pressure drift of the secondary standard~\footnote{When SiN~3 is exposed to uncontrolled environmental changes (\textit{e.g.} long-term exposure to atmosphere), the drift in \(\Gamma_0\) is no longer consistent with zero, but it remains equivalent to a pressure drift more than \(10\) times lower than the secondary standard}.
Together, the Allan deviation and baseline drift measurements illustrate the potential for high stability pressure sensing with optomechanical devices.


\subsection{\label{sec:den_meas} Density Measurement}


To validate the density characterization of our devices in Appendix~\ref{sec:characterization}, we use the mechanical damping measurements of Sec.~\ref{sec:prim_op} to infer the thin-film density of each device.
To perform the density measurement, we compute the characteristic acceleration
\begin{equation}
    \label{eq:char_acc}
    a_c = \frac{\Gamma'_t}{1+\pi/4}\sqrt{\frac{\pi k_B T}{8 m_g}}
\end{equation}
from the measured \(\Gamma'_t\).
We then fit the characteristic acceleration data for each \{sensor, test gas\} pair to
\begin{equation}
    \label{eq:fitfunc}
    a_c=S_m P'_{ref}+b,
\end{equation}
where the measured sensitivity \(S_m\) and intercept \(b\) are fit parameters.
Combining \(S_m\) with the ellipsometer measurements of \(h\) yields the \textit{in situ} density of our devices \(\rho_m=1/S_m h\).
We provide details of the fitting procedure and density uncertainty analysis in Appendix~\ref{sec:fitting}.

\begin{table}
    \caption{Thin-film density for each device determined via prior characterization (\(\rho\)) or \textit{in situ} measurement (\(\rho_m\)).
    Parenthetical quantities represent total (statistical and non-statistical) standard uncertainties at \(k=1\).}
    \label{tab:comparison}
    \begin{ruledtabular}
        \begin{tabular}{cccc}
            Device\T & Gas & \(\rho~(\si[per-mode=symbol]{\kilo\gram\per\cubic\meter})\) & \(\rho_m~(\si[per-mode=symbol]{\kilo\gram\per\cubic\meter})\)\B \\
            \hline
            SiN~1\T & Ar & \(2810(141)\) & \(2822(15)\) \\
            & He & \(2810(141)\) & \(2719(11)\) \\
            & N\(_2\) & \(2810(141)\) & \(2822(12)\)\B \\
            \hline
            SiN~3\T & Ar & \(2810(141)\) & \(2818(28)\) \\
            & He & \(2810(141)\) & \(2761(29)\) \\
            & N\(_2\) & \(2810(141)\) & \(2870(27)\)\B \\
            \hline
            SiC~2\T & Ar & \(3210(20)\) & \(3233(29)\) \\
            & He & \(3210(20)\) & \(3127(28)\) \\
            & N\(_2\) & \(3210(20)\) & \(3219(29)\)\B \\
        \end{tabular}
    \end{ruledtabular}
\end{table}



We compare \(\rho_m\) to \(\rho\) for each device in Table~\ref{tab:comparison}.
All \(\rho_m\) agree with \(\rho\) at two standard uncertainties (\(k=2\)), except the \{SiC~2, He\} combination, which agrees at three standard uncertainties (\(k=3\)).
Additionally, the densities measured with Ar and N\(_2\) for each device agree with each other at two standard uncertainties (\(k=2\)).
Together, the two results above confirm that our sensors are accurate to within the uncertainties quoted in the discussion of Fig.~\ref{fig:pvp} for heavy gases.
The total fractional uncertainty in \(\rho_m\) for silicon nitride devices is \(1~\si{\percent}\) or less.
Our method of measuring the mechanical damping as a function of pressure can complement other techniques for characterizing the density of SiN thin films, such as Rutherford backscattering~\cite{markwitz1993, huszank2016} or mechanical resonance methods~\cite{Xu2023}.

\section{\label{sec:conclusion}Conclusion}

We have carefully explored the performance and limitations of optomechanical pressure sensors.
By comparing damping-based pressure measurements from three optomechanical devices to a secondary standard, we find that our optomechanical pressure sensors are consistent with NIST's primary pressure standards when measuring heavy gases.
This is a necessary step towards using optomechanical pressure sensors as primary pressure sensors in industrial or metrological applications.
The agreement between our sensors and NIST's standards does not depend on the sensor material or oscillator geometry.
Owing to its fabrication from a single-crystal material, the SiC sensor exhibits a total uncertainty of \(1.1~\si{\percent}\).
The uncertainty of our secondary standard allows us to determine the thin-film density of our SiN sensors with \(1~\si{\percent}\) or better total uncertainty.
Precise measurements of optomechanical damping as a function of pressure can characterize the density of other optomechanical sensors, which must often have a well-known mass to achieve accurate measurements.


Our optomechanical sensors possess exceptional short-term and long-term stability.
The Allan deviation of one SiN trampoline indicates that optomechanical sensors can reach precision better than \(1~\si{\percent}\).
The long-term baseline drift of our SiN trampolines is \(10\) times better than that of our secondary standard.
The stability measurements suggest that the useful operating range of our SiN trampolines can be extended down to pressures on the order of \(100~\si{\micro\pascal}\).
Together, our results indicate that optomechanical pressure sensors can meet or exceed the performance of capacitance diaphragm and spinning rotor gauges.


\section*{Acknowledgements}

We thank E. Norrgard and A. Chijioke for their careful reading of the manuscript.
We also thank S. Eckel, J. Prothero, and J. Ricker for experimental assistance, as well as A. Migdall and I. Spielman for loaning us equipment used in this work.

\appendix
\section{\label{sec:devices}Trampoline Fabrication}

The trampoline membranes were fabricated using a two-sided process.
Silicon nitride was deposited on the silicon substrate (\(525~\si{\micro\meter}\) thick) using low-pressure chemical vapor deposition (LPCVD).
The trampolines were patterned on the front side using direct laser writing lithography and transferred into the silicon nitride through reactive ion etching (RIE).
A low-temperature silicon oxide (LTO) film was then deposited as a protection layer for the subsequent steps.
In order to fully release the silicon nitride trampoline membrane, the backside of the silicon wafer was patterned with square openings using direct laser writing lithography.
A combination of RIE and deep reactive ion etching (DRIE) was used to etch the pattern into the silicon substrate to within \(40~\si{\micro\meter}\) of the front surface.
After the substrate was diced and cleaned, the trampoline membranes were fully released using a timed chip-by-chip etch in KOH.
Finally, each chip is dipped in a buffered oxide etch to remove the LTO protection layer.

\section{\label{sec:characterization}Density and Thickness Characterization}

We determine the device thickness \(h\) using ellipsometry.
We perform ellipsometer measurements at four distinct points on the device frame.
We take \(h\) as the mean of the four measurements with associated standard uncertainty \(u_h\) given by the standard deviation of the four measurements.

Silicon nitride thin films have lower density than bulk silicon nitride.
We therefore characterize the density of our films by comparing the mass of a single-side-polished silicon wafer before and after deposition of a \(1~\si{\micro\meter}\) thick SiN layer using a calibrated scale.
As a result of using LPCVD, SiN is deposited on both sides of the wafer.
The thickness of the SiN layer on the polished side was measured using an ellipsometer and the thickness variation across the wafer was included in the density calculation.
Our calculation assumes that the silicon nitride layer on the unpolished side of the wafer has the same thickness and thickness variation as the layer on the polished side.
We estimate that the relative standard uncertainty of our device density \(u_\rho/\rho\) is \(5~\si{\percent}\), to account for potential density variation with layer thickness and from batch-to-batch.
The nominal density of SiN~1 and SiN~3 is \(\rho=2810(141)~\si{\kilo\gram\per\cubic\meter}\), which is consistent with other typically reported values for SiN thin film density at two standard uncertainties (\(k=2\))~\cite{markwitz1993,huszank2016, zotero-2534}.

The membrane of SiC~2 is thin-film, single-crystal 3C-SiC.
Thin-film 3C-SiC has been shown to have the same density as bulk, single-crystal silicon carbide to within the thin-film density measurement uncertainty~\cite{pickering1990,goela1991}.
We therefore take \(\rho=3210(20)~\si{\kilo\gram\per\cubic\meter}\), where the standard uncertainty is the average uncertainty of the thin-film 3C-SiC density measurements reported in Ref.~\cite{pickering1990}.

\section{\label{sec:eq1_der}Damping Rate Derivation}

The dynamics of thin optomechanical membranes and trampolines are described by a damped, 2-dimensional wave equation.
We take the displacement of the mechanical mode \(w(x, y, t)\) aligned with the \(z\) axis, so that the mechanical oscillator lies in the \(xy\) plane.
The damped wave equation is then
\begin{equation}
    \label{eq:2d_wave}
     \rho\frac{\partial^2 w}{\partial t^2}+b \frac{\partial w}{\partial t} = -\sigma\Big(\frac{\partial w}{\partial x^2}+\frac{\partial w}{\partial y^2}\Big),
\end{equation}
where \(\sigma\) is the stress, \(b\) is the damping per unit volume, and \(\rho\) is the density.
Whether Eq.~\eqref{eq:2d_wave} describes a membrane or trampoline depends on the boundary conditions.
The boundary conditions for a square membrane are \(w(0, y, t)=w(L, y, t)=w(x, 0, t)=w(x, L, t)=0\), where \(L\) is the edge length of the membrane, while trampolines have more complicated boundary conditions.

The damped wave equation is solved via separation of variables, where we take \(w(x, y, t) = a_{nm}(t)\psi_{n}(x)\phi_m(y)\) with \(n\) and \(m\) integers that index the mechanical mode.
The resulting equations for \(\psi_n\), \(\phi_m\), and \(a_{nm}\) are
\begin{equation}
    \label{eq:2d_sepvar}
    \begin{split}
        \frac{\partial^2\psi_n}{\partial x^2} & = -\lambda_n^2 \psi_n, \\
        \frac{\partial^2\phi_m}{\partial y^2} & = -\lambda_m^2 \phi_m, \\
        \rho\frac{\partial^2 a_{nm}}{\partial t^2} & = -b\frac{\partial a_{nm}}{\partial t}-\rho\omega^2 a_{nm},
    \end{split}
\end{equation}
where \(\omega=\sqrt{(\lambda_n^2+\lambda_m^2)\sigma/\rho}\).
We follow the convention of Ref.~\cite{Hauer2013} where \(a_{nm}\) has units of length while \(\psi_n\) and \(\phi_m\) are unitless.
For a square membrane, we then have \(\psi_n = \sin(n\pi x/L)\), \(\phi_m = \sin(m\pi y/L)\), \(\lambda_n=n\pi/L\), and \(\lambda_m=m\pi/L\).

We wish to determine the mechanical damping rate of \(a_{nm}\) in the molecular flow regime.
We make the approximation that all other damping sources are negligible.
The top and bottom surfaces of the oscillator each have area \(A_\perp\) and any sidewalls -- due to, for example, holes cut in a membrane to form a photonic crystal mirror -- have total area \(A_\parallel\).
Because gas damping occurs only at the surface of the oscillator, the damping rate per unit volume is then
\begin{equation}
    \label{eq:vol_damp}
    b = \frac{d\beta_\perp}{dA_\perp}\big(\delta(z-h/2){+}\delta(z+h/2)\big){+}\frac{d\beta_\parallel}{dA_\parallel}\delta(f(x,y)),
\end{equation}
where \(\delta(z)\) is the Dirac delta function, \(\frac{d\beta_\perp}{dA_\perp}\) is the damping coefficient per unit area of the top or bottom surface, \(\frac{d\beta_\parallel}{dA_\parallel}\) is the damping per unit area of any sidewalls, and the function \(f(x,y)=0\) when \((x,y) \in A_\parallel\).
From Refs.~\cite{Cavalleri2010, Martinetz2018, Christian1966}, we have
\begin{equation}
    \label{eq:area_damp}
    \begin{split}
        \frac{d\beta_\perp}{dA_\perp} & = P\sqrt{\frac{2m_g}{\pi k_B T}}\Big((1+\pi/4)\alpha+2 (1-\alpha)\Big),\\
        \frac{d\beta_\parallel}{dA_\parallel} & = P\alpha\sqrt{\frac{m_g}{2\pi k_B T}}.\\
    \end{split}
\end{equation}

We first consider the case of a square membrane.
Membranes have no sidewalls exposed to the surrounding gas, so \(A_\parallel=0\) and therefore \(\frac{d\beta_\parallel}{dA_\parallel}\) does not contribute to the mechanical damping.
We insert Eq.~\eqref{eq:vol_damp} into the last line of Eqns.~\ref{eq:2d_sepvar} and integrate over \(z\) to find
\begin{equation}
    \label{eq:damp_mem}
    \rho h\frac{\partial^2 a_{nm}}{\partial t^2} +2\frac{d\beta_\perp}{dA_\perp}\frac{\partial a_{nm}}{\partial t}+\rho h\omega^2 a_{nm} = 0.
\end{equation}
After dividing Eq.~\eqref{eq:damp_mem} by \(\rho h\), we identify the mechanical damping rate for \(a_{nm}\) as
\begin{equation}
    \Gamma_m = \frac{2}{\rho h}\frac{d\beta_\perp}{dA_\perp} = \frac{(1+\pi/4)\alpha+2(1-\alpha)}{\rho h}\sqrt{\frac{8 m_g}{\pi k_B T}} P,
\end{equation}
which is Eq.~\eqref{eq:mol_flow}.

Deriving the gas-induced mechanical damping rate for a trampoline, or a device with a photonic crystal mirror, is more complicated.
Because a trampoline has sidewalls (\textit{i.e.}, non-trivial boundary conditions), the damping coefficient per unit volume \(b\) becomes a function of \(x\) and \(y\) (see Eq.~\eqref{eq:vol_damp}).
Equation~\eqref{eq:2d_wave} is thus no longer separable. 
However, in the limit that \(\sigma\) is significantly larger than \(h\, d\beta_\parallel/dA_\parallel\), 
it is justifiable to approximate the sidewall damping as uniformly distributed across the area of the device, which results in
\begin{equation}
    \label{eq:vol_damp2}
    b = \Big(\frac{d\beta_\perp}{dA_\perp}{+}\frac{d\beta_\parallel}{dA_\parallel}\frac{A_\parallel}{2 A_\perp}\Big)\big(\delta(z-h/2){+}\delta(z+h/2)\big).
\end{equation}
Integrating the last line of Eqns.~\eqref{eq:2d_sepvar} over \(z\) then yields
\begin{equation}
    \label{eq:damp_tramp}
    \rho h\frac{\partial^2 a_{nm}}{\partial t^2} +\Big(2\frac{d\beta_\perp}{dA_\perp}{+}\frac{d\beta_\parallel}{dA_\parallel}\frac{A_\parallel}{A_\perp}\Big)\frac{\partial a_{nm}}{\partial t}+\rho h\omega^2 a_{nm} = 0.
\end{equation}
By dividing Eq.~\eqref{eq:damp_tramp} by \(\rho h\), we see that the damping rate for a trampoline or device with a photonic crystal mirror is given by Eq.~\eqref{eq:mol_flow} with an approximate correction
\begin{equation}
    \label{eq:tramp_corr}
    \delta\Gamma_m \approx \frac{d\beta_\parallel}{dA_\parallel}\frac{A_\parallel}{\rho h A_\perp} = \frac{\alpha}{\rho h}\frac{A_\parallel}{4 A_\perp}\sqrt{\frac{8 m_g}{\pi k_B T}} P,
\end{equation}
which is small for thin oscillators (\(A_{\parallel}\ll A_{\perp}\)).
The fractional correction \(\delta\Gamma_m/\Gamma_m \le A_{\parallel}/(4 (1+\pi/4) A_{\perp})\) with the equality occurring in the diffuse reflection limit \(\alpha\rightarrow 1\).


For the trampoline devices SiN~1 and SiN~3, we estimate the additional sidewall damping \(\delta\Gamma_m\) from Eq.~\eqref{eq:tramp_corr}. We use the designed perimeter of the top surface and the measured device thickness \(h\), which we multiply to find \(A_\parallel\), as well as the designed top surface area \(A_\perp\).
For SiN~1, we also include the reduction of \(A_\perp\) and increase in \(A_\parallel\) due to the hole pattern that creates the photonic crystal mirror.
The fractional increase in damping \(\delta\Gamma_m/\Gamma_m\) when \(\alpha=1\) is \(0.2~\si{\percent}\) for SiN~1 and \(0.3~\si{\percent}\) for SiN~3.
The damping correction for SiN~3 is slightly larger than the correction for SiN~1 because its long tethers contribute more sidewall damping than the photonic crystal patterned into SiN~1.
If SiN~1 did not have a photonic crystal mirror, it would have \(\delta\Gamma_m/\Gamma_m\approx 0.1~\si{\percent}\).
For both devices, \(\delta\Gamma_m/\Gamma_m \ll u_\rho/\rho\), so the the extra sidewall damping \(\delta\Gamma_m\) is negligible in our pressure measurements.

\section{\label{sec:fitting}Detailed Fitting Procedure}

We perform the linear fit to the characteristic acceleration data using maximum likelihood estimation on the orthogonal distance.
When parameterized by the slope \(S_m\) and intercept \(b\), the orthogonal distance of the \(i\)th data point \((P'_{{\rm ref}, i}, a_{c,i})\) from a line is \(d_i = a_{c, i} - S_m P'_{{\rm ref}, i} - b\).
The log-likelihood for the orthogonal distance is
\begin{equation}
    \label{eq:loglike}
    \begin{split}
        \ln \mathcal{L} & = \frac{1}{2}\ln\Big(\mathrm{det}\big((C(\vec{a}_c)+S^2_m C(\vec{P}'_{\rm ref}))^{-1}\big)\Big) \\
        & + \frac{1}{2}\vec{d}^{\;\rm T}(C(\vec{a}_c)+S^2_m C(\vec{P}'_{\rm ref}))^{-1}\vec{d},
    \end{split}
\end{equation}
where \(C(\vec{a}_c)\) and \(C(\vec{P}'_{\rm ref})\) are the covariance matrices for the set of measurements \(\vec{a}_c\) and \(\vec{P}'_{\rm ref}\), respectively.
We note that the covariance matrices are not diagonal (see Appendix~\ref{sec:uncertainty}) and that the orthogonal distance regression algorithm accounts for these correlated uncertainties.
To ensure that we are not fitting into the viscous flow regime, we include data up to a variable cut-off pressure \(P'_{\rm ref, \,cut}\) in our minimization of Eq.~\eqref{eq:loglike} and observe that \(S_m\) is independent of \(P'_{\rm ref, \,cut}\) for \(P'_{\rm ref, \,cut}\lesssim 100~\si{\pascal}\).
The maximum likelihood fit results with \(P'_{\rm ref, \,cut}= 10~\si{\pascal}\), chosen well below the onset of viscous flow, are used to calculate \(\rho_m=1/S_m h\) in Sec.~\ref{sec:den_meas}.

We assess the fit quality using \(\chi^2_\nu = \chi^2/\nu\), where \(\nu\) is the number of degrees of freedom in the fit and we include the correlated uncertainties in our calculation of \(\chi^2\) following Ref.~\cite{schmelling1995}.
Table~\ref{tab:chisq} shows both \(\chi^2_\nu\) and \(\nu\) for each combination of sensor and test gas.
All fits pass the \(\chi^2\) test, where the probability of observing a \(\chi^2_\nu\) at least as large as those in Table~\ref{tab:chisq} exceeds \(5~\si{\percent}\)~\cite{Bevington1992}.

\begin{table}
    \caption{Measured sensitivity \(S_m\), reduced \(\chi^2\) statistic \(\chi^2_\nu\) and number of degrees of freedom \(\nu\) for the linear fit to each device and gas (see text).
    Parenthetical quantities represent total (statistical and non-statistical) standard uncertainties at \(k=1\).}
    \label{tab:chisq}
    \begin{ruledtabular}
        \begin{tabular}{ccccc}
            Device\T & Gas & \(S_m~(\si[per-mode=symbol]{\square\meter\per\kilo\gram})\) & \(\chi^2_\nu\) & \(\nu\)\B \\
            \hline
            SiN~1\T & Ar & \(1608(7)\) & \(0.40\) & \(30\) \\
            & He & \(1669(5)\) & \(0.68\) & \(43\) \\
            & N\(_2\) & \(1608(5)\) & \(0.51\) & \(32\)\B \\
            \hline
            SiN~3\T & Ar & \(1583(12)\) & \(0.26\) & \(21\) \\
            & He & \(1616(13)\) & \(0.97\) & \(23\) \\
            & N\(_2\) & \(1554(11)\) & \(0.60\) & \(20\)\B \\
            \hline
            SiC~2\T & Ar & \(6030(11)\) & \(0.84\) & \(25\) \\
            & He & \(6237(13)\) & \(0.86\) & \(26\) \\
            & N\(_2\) & \(6058(11)\) & \(0.86\) & \(26\)\B \\
        \end{tabular}
    \end{ruledtabular}
\end{table}

We determine the uncertainty in the fit parameters \(S_m\) and \(b\) using a Markov chain Monte Carlo (MCMC) method.
Markov chain Monte Carlo, which we implement using the {\tt emcee} Python package~\cite{foreman-mackey2013}, allows our parameter uncertainty estimation to include the correlated uncertainties in our measurements (see Appendix~\ref{sec:uncertainty}).
Table~\ref{tab:chisq} shows \(S_m\) and its total standard uncertainty for each combination of sensor and test gas.
The fitted intercept \(b\) agrees with \(0~\si[per-mode=symbol]{\meter\per\square\second}\) within two standard uncertainties (\(k=2\)) for all devices and gases, except for SiN~1 with Ar, where it agrees within three standard uncertainties (\(k=3\)).
We compute the \textit{in situ} density uncertainty \(u_{\rho_m}\) by propagating the uncertainty in \(S_m\) and \(h\).
Because we measure \(S_m\) under the assumption that the mechanical damping rate is given by Eq.~\eqref{eq:mol_flow}, we take the approximate sidewall damping correction \(\delta\Gamma_m\) derived in Appendix~\ref{sec:eq1_der} as an additional source of uncertainty for SiN~1 and SiN~3.
The calculated \(u_{\rho_m}\) are reported in Table~\ref{tab:comparison}.

\section{\label{sec:uncertainty}Uncertainty Analysis}

The uncertainties reported in Sec.~\ref{sec:res} are computed using the full covariance matrix for both the secondary standard and optomechanical sensor.
We perform our analysis using the covariance matrix because many of the non-statistical uncertainties of the secondary standard are correlated.
The total uncertainty in a set of measurements \(\vec{X}\) of a physical quantity \(X\) is described by the covariance matrix \(C(\vec{X})\) with elements
\begin{equation}
    \label{eq:tot_cov}
    C(X_i, X_j) = \sum_{Y} C_{Y}(X_i, X_j),
\end{equation}
where integers \(i\) and \(j\) index the measurements of \(X\), \(Y\) labels a source of uncertainty, and we take uncertainties due to distinct sources \(Y\) to be uncorrelated, which is the case for the significant uncertainty sources in our measurements.
The uncertainty in the set of measurements \(\vec{X}\) due to source \(Y\) is given by
\begin{equation}
    \label{eq:par_cov}
    C_{Y}(X_i, X_j) = u_{Y}(X_i)u_{Y}(X_j)r_{Y}(X_i, X_j),
\end{equation}
where \(u_{Y}(X_i)\) is the standard uncertainty (\(k=1\)) in the \(i\)th measurement of \(X\) (\textit{i.e.}, \(X_i\)) due to source \(Y\) and \(r_{Y}(X_i, X_j)\) is the correlation coefficient between the \(i\)th and \(j\)th measurements.
The main results of Sec.~\ref{sec:res} use background subtraction, which introduces additional correlations into the covariance matrices.
Specifically, the elements of the covariance matrix for \(\vec{X}'=\vec{X}-X_0\) due to source \(Y\) are
\begin{equation}
    \label{eq:bg_par_cov}
    \begin{split}
    C_{Y}(X'_i, X'_j) = & \;C_{Y}(X_i, X_j)+\big(u_{Y}(X_0)\big)^2 \\
    &-C_{Y}(X_i, X_0)-C_{Y}(X_0, X_j).
    \end{split}
\end{equation}
Importantly, background subtraction significantly suppresses highly correlated sources of uncertainty in the secondary standard.

\begin{figure*}
    \centering
    \includegraphics[width=\textwidth]{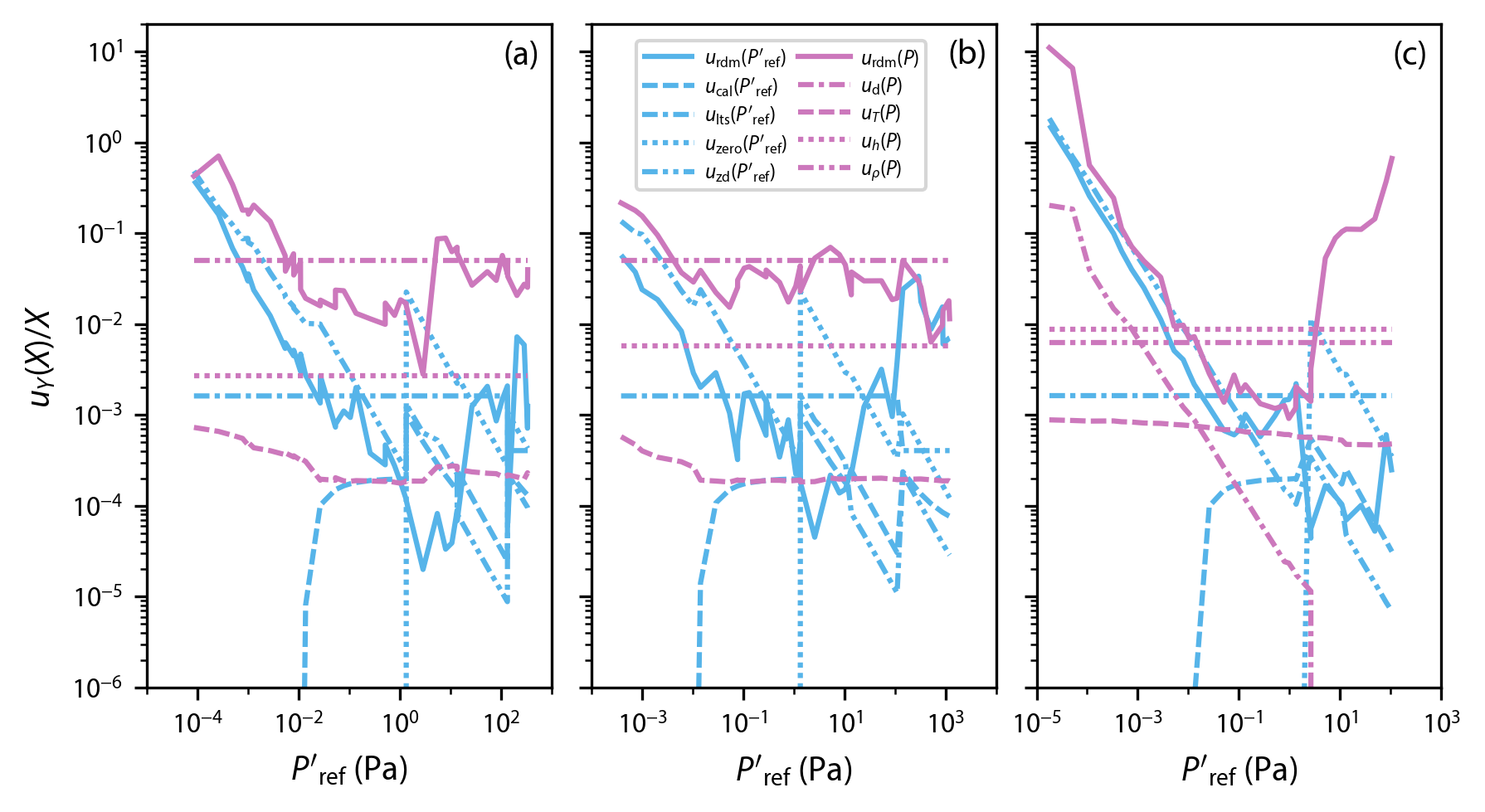}
    \caption{Example uncertainty budget for Ar test gas measurements with SiN~1 (a), SiN~3 (b), and SiC~2 (c).
    Blue and purple colored curves show the fractional standard uncertainty due to each source affecting the secondary standard and optomechanical sensor, respectively.
    The standard uncertainties are given by the diagonal elements of the background-subtracted covariance matrices for each uncertainty component (see text).}
    \label{fig:characcvp_unc}
\end{figure*}

We show example uncertainty budgets for the Ar test gas measurements of each device in Fig.~\ref{fig:characcvp_unc}.
The only difference between the uncertainty budget for the direct pressure comparison (\textit{i.e.}, \(P\) vs. \(P'_{\rm ref}\), see Fig.~\ref{fig:pvp}) and for the density measurement (\textit{i.e.}, \(a_c\) vs. \(P'_{\rm ref}\), see Sec.~\ref{sec:den_meas}) is that uncertainty due to \(\rho\) does not contribute for the density measurement error budget (see Eq.~\eqref{eq:char_acc}).
Because the fractional uncertainties in \(P\) and \(a_c\) due to source \(Y\) are the same, we choose to label all uncertainties of the optomechanical devices with physical quantity \(P\), rather than physical quantity \(a_c\).
Blue and purple lines in Fig.~\ref{fig:characcvp_unc} show the fractional standard uncertainty at \(k=1\) associated with each uncertainty source for the secondary standard and optomechanical sensor, respectively.
The fractional standard uncertainties are computed from the diagonal elements of the background-subtracted covariance matrix (see Eq.~\ref{eq:bg_par_cov}) as 
\begin{equation}
    \frac{u_{Y}(X'_i)}{X'_{i}} = \frac{\sqrt{(u_{Y}(X_i))^2+(u_{Y}(X_0))^2-2 C_{Y}(X_i, X_0)}}{X'_{i}}.
\end{equation}
We detail the construction of the covariance matrix, Eq.~\eqref{eq:par_cov}, for each uncertainty component below.

\subsection{\label{sec:unc_optomech}Optomechanical Sensor}

The \(P\) and \(a_c\) measurements of the optomechanical sensors have eight and six sources of uncertainty, respectively.
The uncertainty sources common to both measurements are the statistical uncertainty in \(\Gamma_t\) (and \(\Gamma_0\)), long-term drift uncertainty in \(\Gamma_0\), uncertainty in the gas temperature \(T\), uncertainty in the gas composition, uncertainty in the digital oscilloscope timebase, and uncertainty in the linearity of the Michelson interferometer.
As we explore below, the last three sources of uncertainty are negligible compared to the first three sources.
Additionally, the pressure measurements must include uncertainty in the sensor density \(\rho\) and thickness \(h\).

First, the statistical uncertainty in \(P\) is given by
\begin{equation}
    \label{eq:gam_rdm}
    C_{\rm rdm}(P_{i}, P_{j}) = \frac{\partial P_i}{\partial \Gamma_{t, i}} u_{\rm rdm}(\Gamma_{t,i}) \frac{\partial P_j}{\partial \Gamma_{t, j}} u_{\rm rdm}(\Gamma_{t,j})\delta_{ij},
\end{equation}
where \(\delta_{ij}\) is the Kronecker delta.
The uncertainty in individual exponential fits (for ring-down measurements) or Lorentzian fits (for thermo-mechanical noise measurements) is negligible compared to the spread of repeated \(\Gamma_t\) measurements.
We therefore take the standard deviation of five repeated measurements as \(u_{\rm rdm}(\Gamma_{t,i})\).
We chose to index our measurements such that \(\Gamma_{t,0}\) is measured at base pressure (\textit{i.e.} \(\Gamma_{t,0} \equiv \Gamma_0\)), so Eq.~\eqref{eq:gam_rdm} includes the statistical uncertainty in \(\Gamma_0\).

Second, there is uncertainty in \(P\) due to long-term drift in \(\Gamma_0\).
We measure the base pressure damping rate \(\Gamma_0\) for each device at the beginning of each measurement run.
For devices SiN~1 and SiN~3, the average daily drift in \(\Gamma_0\) is negligible compared the statistical uncertainty of the individual \(\Gamma_0\) measurements (see Fig.~\ref{fig:allan}(b)).
A measurement run takes approximately \(8~\si{\hour}\), so we do not include drift uncertainty in the uncertainty budget SiN~1 or SiN~3.
For SiC~2, we observe an average daily drift in \(\Gamma_0\) of \(B_{\Gamma} = 0.03~\si{\per\second}\text{/d}\), which is, relative to the statistical uncertainty, significant for ring-down data and negligible for thermo-mechanical noise data.
The covariance matrix for the drift is
\begin{equation}
    \begin{split}
        C_{\rm d}(P_{i},P_{j}) & = \frac{\partial P_i}{\partial \Gamma_{t, i}} u_{\rm d}(\Gamma_{t,i}) \frac{\partial P_j}{\partial \Gamma_{t, j}} u_{\rm d}(\Gamma_{t,j})\delta_{ij} \\
        & = \frac{\partial P_i}{\partial \Gamma_{t, i}} B_{\Gamma}(t_i-t_0) \frac{\partial P_j}{\partial \Gamma_{t, j}}B_{\Gamma}(t_j-t_0)\delta_{ij},
    \end{split}
\end{equation}
where \(t_i\) is the time of the \(i\)th measurement and \(t_0\) is the time of the first measurement of the run.

Third, there is uncertainty in the gas temperature.
The two PRTs monitoring the vacuum chamber temperature report an approximately \(1~\si{\kelvin}\) temperature gradient across the chamber.
The gradient decreases with gas pressure, presumably due to the thermal conductivity of the gas, to approximately \(200~\si{\milli\kelvin}\) at \(1000~\si{\pascal}\).
The PRTs have a calibration uncertainty of approximately \(1~\si{\milli\kelvin}\), which is negligible compared to the gradient.
We therefore take the temperature gradient as the standard uncertainty in the gas temperature with covariance matrix
\begin{equation}
    \begin{split}
        C_{T}(P_{i}, P_{j}) & = u_{T}(P_{i})u_{T}(P_{j})\delta_{ij} \\
        & = \frac{\partial P_i}{\partial T_{i}}\Delta T_i \frac{\partial P_j}{\partial T_{j}} \Delta T_j\delta_{ij},
    \end{split}
\end{equation}
where \(\Delta T_i\) is the temperature gradient during the \(i\)th measurement.

Fourth, there is uncertainty in the gas composition. Strictly, the mass that enters Eq.~\eqref{eq:meas_p} is the average molecular mass \(m_{\rm avg}=\sum_g P_g m_g/\sum_g P_g\), where \(P_g\) is the partial pressure of gas \(g\) and the sum runs over all gases in the vacuum chamber~\cite{Blakemore2020, barker2024}.
Background subtraction removes uncertainty due to the residual gases at base pressure, whose partial pressures do not vary during a measurement run.
We have verified that there are no leaks in the gas feedline and that its outgassing is insignificant using a residual gas analyzer.
Because we use ultra-high purity test gases, any mass uncertainty will contribute negligibly to the uncertainty budget.

Fifth, there is uncertainty in the timebase of the digital oscilloscope.
Given the age and specifications of the oscilloscope, the timebase uncertainty is at most \(10~\si[per-mode=symbol]{\micro\second\per\second}\).
When propagated through \(\Gamma_t\) to \(P\), the timebase uncertainty then causes \(10~\si[per-mode=symbol]{\micro\pascal\per\pascal}\) fractional pressure uncertainty, which is negligible (see Fig.~\ref{fig:characcvp_unc}).

Sixth, there is uncertainty in the linearity of the Michelson interferometer.
We stabilize the interferometer so that its intensity response to small displacements of the optomechanical sensor is approximately linear, but small deviations from linearity could distort exponential mechanical ring-downs.
Given the amplitude of the interferometer's full sinusoidal intensity response to displacement and our ring-down excitation amplitude, we estimate that the non-linearity is less than \(0.17~\si{\percent}\).
To ensure that such a small non-linearity is not biasing our ring-down measurements, we introduce a variable photodiode voltage cutoff and only perform the ring-down exponential fit on data lying below the cutoff.
We observe no systematic shift in \(\Gamma_t\) as a function of the voltage cutoff, indicating that interferometer non-linearity does not bias our measurements toward lower damping rates.
(The amplitude of the thermo-mechanical noise is significantly lower than that of the ring-down excitation, so any non-linearity is insignificant).

Finally, there is uncertainty in \(P\) due to both \(\rho\) and \(h\).
The covariance matrices for these two sources of uncertainty are
\begin{equation}
    C_{\rho}(P_{i}, P_{j}) = \frac{\partial P_i}{\partial \rho} u_{\rho} \frac{\partial P_j}{\partial \rho} u_{\rho},
\end{equation}
and
\begin{equation}
    C_{h}(P_{i}, P_{j}) = \frac{\partial P_i}{\partial h} u_{h} \frac{\partial P_j}{\partial h} u_{h},
\end{equation}
respectively.
We note that \(C_{\rho}(P_{i}, P_{j})\) and \(C_{h}(P_{i}, P_{j})\) do not depend on \(i\) or \(j\), so they are perfectly correlated over \(P\).

\subsection{\label{sec:unc_cdg}Secondary Standard}

The \(P_{\rm ref}\) measurements of the secondary standard have seven sources of uncertainty.
Five uncertainty sources contribute non-negligibly to the total uncertainty of our measurements: statistical uncertainty, uncertainty of the initial calibration, uncertainty due to long-term calibration drift, uncertainty in determining the zero pressure reading, and uncertainty due to drift in the zero pressure reading.
The latter two uncertainty sources arise because CDGs are fundamentally differential gauges that employ a passively pumped reference vacuum on one side of the gauge diaphragm to measure absolute pressure on the other side.
Mechanical aging and temperature changes then lead to variation in the CDG ``zero'', which is the indicated pressure when the absolute pressure is below the gauge's resolution~\cite{hyland1985,miller1997}.
There are two sources of uncertainty that do not contribute significantly to our measurements: uncertainty in the calibration of the digital multimeter (DMM) that records CDG measurements, and uncertainty in the thermal transpiration correction to the CDG readings.
We describe our construction of the covariance matrix for each uncertainty source below.

First, the random statistical uncertainty has
\begin{equation}
    C_{\rm rdm}(P_{{\rm ref},i},P_{{\rm ref},j}) = u_{\rm rdm}(P_{{\rm ref},i})u_{\rm rdm}(P_{{\rm ref},j})\delta_{ij},
\end{equation}
where we take \(u_{\rm rdm}(P_{{\rm ref},i})\) to be the standard deviation of \(10\) repeated measurements and \(\delta_{ij}\) is the Kronecker delta.

Second, we consider the uncertainty in the initial calibration of each CDG within the secondary standard by the NIST primary standards.
Because each CDG has three independently calibrated gain settings, the calibration uncertainty is perfectly correlated for each \{CDG, gain\} pair (and uncorrelated between pairs), so 
\begin{equation}
    C_{\rm cal}(P_{{\rm ref},i},P_{{\rm ref},j}) = u_{\rm cal}(P_{{\rm ref},i})u_{\rm cal}(P_{{\rm ref},j})\delta_{k_i k_j}\delta_{l_i l_j},
\end{equation}
where \(k_i\) (\(l_i\)) indexes the CDG (gain setting) for \(P_{{\rm ref},i}\).
The secondary standard's calibration report provides \(u_{\rm cal}(P_{\rm ref})\) for each CDG and gain setting.
For a given CDG and gain setting, \(u_{\rm cal}(P_{\rm ref})\) is approximately constant over the calibration pressure range and we account for its small increase with \(P_{\rm ref}\) with linear interpolation.

Third, we describe the uncertainty associated with long-term drift of the secondary standard calibration.
The calibration report provides the average fractional calibration shift between successive calibrations for each CDG in the secondary standard.
The reported calibration shift is averaged over gain settings, so the covariance matrix for the long-term stability uncertainty is
\begin{equation}
    \begin{split}
        C_{\rm lts}(P_{{\rm ref},i},P_{{\rm ref},j}) & = u_{\rm lts}(P_{{\rm ref},i})u_{\rm lts}(P_{{\rm ref},j})\delta_{k_i k_j} \\
        & = A_{k_i}P_{{\rm ref}, i}\,A_{k_j}\,P_{{\rm ref}, j}\,\delta_{k_i k_j},
    \end{split}
\end{equation}
where \(A_{k_i}\) is the fractional calibration shift coefficient for CDG \(k_i\), averaged over the \(27\)~year calibration history of the secondary standard.

Fourth, we include uncertainty in the initial determination of each CDG zero.
We zero each \{CDG, gain\} pair by recording the reading at the chamber base pressure before beginning a pressure measurement run.
We use the average of \(10\) repeated measurements as the zero reading, which is subtracted from all subsequent pressure readings (until the secondary standard is zeroed again before the next measurement run).
We take the standard uncertainty of the zero \(u_{\rm zero}(k_i, l_i)\) to be the standard deviation of the \(10\) measurements.
The covariance matrix for the initial zero uncertainty is then
\begin{equation}
    \begin{split}
        C_{\rm zero}(P_{{\rm ref},i},P_{{\rm ref},j}) & = u_{\rm zero}(P_{{\rm ref},i})u_{\rm zero}(P_{{\rm ref},j})\delta_{k_i k_j}\delta_{l_i l_j} \\
        & = u_{\rm zero}(k_i, l_i)u_{\rm zero}(k_j, l_j)\delta_{k_i k_j}\delta_{l_i l_j}.
    \end{split}
\end{equation}

Fifth, there is drift uncertainty due to the CDG zero changing with time.
To assess the zero drift uncertainty, we log the average zero readings each time that we zero the secondary standard and compute the daily zero drift coefficient \(B_{kl}\) for CDG \(k\) and gain setting \(l\), averaged across our measurement campaign.
For our tightly temperature-controlled secondary standard, the zero drift is approximately monotonic with small discontinuous jumps~\cite{miller1997, hyland1985}, so we take the covariance matrix for the zero drift as
\begin{equation}
    \label{eq:cov_zd}
    \begin{split}
        C_{\rm zd}(P_{{\rm ref},i},P_{{\rm ref},j}) & = u_{\rm zd}(P_{{\rm ref},i})u_{\rm zd}(P_{{\rm ref},j})\delta_{ij} \\
        & = B_{k_i l_i}(t_i-t_z)B_{k_j l_j}(t_j-t_z)\delta_{ij},
    \end{split}
\end{equation}
where \(t_i\) is the time of the \(i\)th measurement and \(t_z\) is the time of the last secondary standard zero.
In principle, the zero drift uncertainty is correlated in time with a correlation envelope that decays with \(|t_i-t_j|\).
However, we are unaware of any systematic studies of the zero drift correlation time of CDGs, so we use the more conservative estimate of Eq.~\eqref{eq:cov_zd}, as suggested by the results of Ref.~\cite{hyland1985, miller1997}.

Sixth, there is uncertainty in the calibration of the DMM that records the secondary standard readings.
We have verified that the DMM meets its \(90\) day accuracy specification using a calibrated voltage standard.
The \(90\) day accuracy specification for our operating voltage range is \(20~\si[per-mode=symbol]{\micro\volt\per\volt}\) of reading.
After propagating the uncertainty to \(P_{\rm ref}\), the calibration uncertainty corresponds to a fractional pressure uncertainty of approximately \(20~\si[per-mode=symbol]{\micro\pascal\per\pascal}\), which is negligible compared to other sources of uncertainty (see Fig.~\ref{fig:characcvp_unc}).
The DMM also exhibits a \(50~\si{\micro\volt}\) uncertainty due to the resolution of its analog-to-digital converter, but this uncertainty is random and thus already contained within the covariance matrix of the statistical uncertainty above.

Seventh, there is uncertainty in the thermal transpiration correction to \(P'_{\rm ref}\).
The secondary standard is connected to the vacuum system by a tube with a \(4.6~\si{\milli\meter}\) inner diameter and it is stabilized to \(23.78(1)~\si{\celsius}\).
The vacuum system has a nominal temperature of \(21~\si{\celsius}\).
We therefore correct \(P'_{\rm ref}\) for the thermal-transpiration-induced pressure gradient between the secondary standard and the optomechanical sensor using the Takaishi-Sensui equation~\cite{ricker2017,takaishi1963,poulter1983}.
The Takaishi-Sensui equation depends on the pressure measured by the secondary standard, the secondary standard temperature, and the vacuum chamber temperature.
In principle, uncertainties in each of the above quantities impact the thermal transpiration correction uncertainty.
However, the measured pressure uncertainty and vacuum chamber temperature uncertainty have already been included in the analysis above and in Appendix~\ref{sec:unc_optomech}, respectively.
Because the uncertainties in these quantities are perfectly self-correlated, we do not include them here to avoid double counting.
The uncertainty in the secondary standard temperature is less than \(1~\si{\milli\kelvin}\) during a \(5\) repeat measurement of \(\Gamma_t\), which corresponds to a fractional thermal transpiration correction uncertainty less than \(2~\si[per-mode=symbol]{\micro\pascal\per\pascal}\).
The thermal transpiration correction uncertainty is thus negligible compared to other sources of uncertainty (see Fig.~\ref{fig:characcvp_unc}).

\bibliography{comparison}

\begin{thebibliography}{46}%
\makeatletter
\providecommand \@ifxundefined [1]{%
 \@ifx{#1\undefined}
}%
\providecommand \@ifnum [1]{%
 \ifnum #1\expandafter \@firstoftwo
 \else \expandafter \@secondoftwo
 \fi
}%
\providecommand \@ifx [1]{%
 \ifx #1\expandafter \@firstoftwo
 \else \expandafter \@secondoftwo
 \fi
}%
\providecommand \natexlab [1]{#1}%
\providecommand \enquote  [1]{``#1''}%
\providecommand \bibnamefont  [1]{#1}%
\providecommand \bibfnamefont [1]{#1}%
\providecommand \citenamefont [1]{#1}%
\providecommand \href@noop [0]{\@secondoftwo}%
\providecommand \href [0]{\begingroup \@sanitize@url \@href}%
\providecommand \@href[1]{\@@startlink{#1}\@@href}%
\providecommand \@@href[1]{\endgroup#1\@@endlink}%
\providecommand \@sanitize@url [0]{\catcode `\\12\catcode `\$12\catcode `\&12\catcode `\#12\catcode `\^12\catcode `\_12\catcode `\%12\relax}%
\providecommand \@@startlink[1]{}%
\providecommand \@@endlink[0]{}%
\providecommand \url  [0]{\begingroup\@sanitize@url \@url }%
\providecommand \@url [1]{\endgroup\@href {#1}{\urlprefix }}%
\providecommand \urlprefix  [0]{URL }%
\providecommand \Eprint [0]{\href }%
\providecommand \doibase [0]{https://doi.org/}%
\providecommand \selectlanguage [0]{\@gobble}%
\providecommand \bibinfo  [0]{\@secondoftwo}%
\providecommand \bibfield  [0]{\@secondoftwo}%
\providecommand \translation [1]{[#1]}%
\providecommand \BibitemOpen [0]{}%
\providecommand \bibitemStop [0]{}%
\providecommand \bibitemNoStop [0]{.\EOS\space}%
\providecommand \EOS [0]{\spacefactor3000\relax}%
\providecommand \BibitemShut  [1]{\csname bibitem#1\endcsname}%
\let\auto@bib@innerbib\@empty
\bibitem [{\citenamefont {Fremerey}(1985)}]{Fremerey1985}%
  \BibitemOpen
  \bibfield  {author} {\bibinfo {author} {\bibfnamefont {J.~K.}\ \bibnamefont {Fremerey}},\ }\bibfield  {title} {\bibinfo {title} {The spinning rotor gauge},\ }\href@noop {} {\bibfield  {journal} {\bibinfo  {journal} {Journal of Vacuum Science and Technology A}\ }\textbf {\bibinfo {volume} {3}},\ \bibinfo {pages} {1715} (\bibinfo {year} {1985})}\BibitemShut {NoStop}%
\bibitem [{\citenamefont {Hyland}\ and\ \citenamefont {Tilford}(1985)}]{hyland1985}%
  \BibitemOpen
  \bibfield  {author} {\bibinfo {author} {\bibfnamefont {R.~W.}\ \bibnamefont {Hyland}}\ and\ \bibinfo {author} {\bibfnamefont {C.~R.}\ \bibnamefont {Tilford}},\ }\bibfield  {title} {\bibinfo {title} {Zero stability and calibration results for a group of capacitance diaphragm gages},\ }\href@noop {} {\bibfield  {journal} {\bibinfo  {journal} {Journal of Vacuum Science \& Technology A}\ }\textbf {\bibinfo {volume} {3}},\ \bibinfo {pages} {1731} (\bibinfo {year} {1985})}\BibitemShut {NoStop}%
\bibitem [{\citenamefont {Salimi}\ \emph {et~al.}(2024)\citenamefont {Salimi}, \citenamefont {Nielsen}, \citenamefont {Pedersen},\ and\ \citenamefont {Dantan}}]{salimi2024}%
  \BibitemOpen
  \bibfield  {author} {\bibinfo {author} {\bibfnamefont {M.}~\bibnamefont {Salimi}}, \bibinfo {author} {\bibfnamefont {R.~V.}\ \bibnamefont {Nielsen}}, \bibinfo {author} {\bibfnamefont {H.~B.}\ \bibnamefont {Pedersen}},\ and\ \bibinfo {author} {\bibfnamefont {A.}~\bibnamefont {Dantan}},\ }\href {https://doi.org/10.48550/arXiv.2312.11915} {\bibinfo {title} {Squeeze film absolute pressure sensors with sub-millipascal sensitivity}} (\bibinfo {year} {2024}),\ \Eprint {https://arxiv.org/abs/2312.11915} {arxiv:2312.11915} \BibitemShut {NoStop}%
\bibitem [{\citenamefont {Magrini}\ \emph {et~al.}(2021)\citenamefont {Magrini}, \citenamefont {Rosenzweig}, \citenamefont {Bach}, \citenamefont {{Deutschmann-Olek}}, \citenamefont {Hofer}, \citenamefont {Hong}, \citenamefont {Kiesel}, \citenamefont {Kugi},\ and\ \citenamefont {Aspelmeyer}}]{Magrini2021}%
  \BibitemOpen
  \bibfield  {author} {\bibinfo {author} {\bibfnamefont {L.}~\bibnamefont {Magrini}}, \bibinfo {author} {\bibfnamefont {P.}~\bibnamefont {Rosenzweig}}, \bibinfo {author} {\bibfnamefont {C.}~\bibnamefont {Bach}}, \bibinfo {author} {\bibfnamefont {A.}~\bibnamefont {{Deutschmann-Olek}}}, \bibinfo {author} {\bibfnamefont {S.~G.}\ \bibnamefont {Hofer}}, \bibinfo {author} {\bibfnamefont {S.}~\bibnamefont {Hong}}, \bibinfo {author} {\bibfnamefont {N.}~\bibnamefont {Kiesel}}, \bibinfo {author} {\bibfnamefont {A.}~\bibnamefont {Kugi}},\ and\ \bibinfo {author} {\bibfnamefont {M.}~\bibnamefont {Aspelmeyer}},\ }\bibfield  {title} {\bibinfo {title} {Real-time optimal quantum control of mechanical motion at room temperature},\ }\href {https://doi.org/10.1038/s41586-021-03602-3} {\bibfield  {journal} {\bibinfo  {journal} {Nature}\ }\textbf {\bibinfo {volume} {595}},\ \bibinfo {pages} {373} (\bibinfo {year} {2021})}\BibitemShut {NoStop}%
\bibitem [{\citenamefont {Barker}\ \emph {et~al.}(2024)\citenamefont {Barker}, \citenamefont {Carney}, \citenamefont {LeBrun}, \citenamefont {Moore},\ and\ \citenamefont {Taylor}}]{barker2024}%
  \BibitemOpen
  \bibfield  {author} {\bibinfo {author} {\bibfnamefont {D.~S.}\ \bibnamefont {Barker}}, \bibinfo {author} {\bibfnamefont {D.}~\bibnamefont {Carney}}, \bibinfo {author} {\bibfnamefont {T.~W.}\ \bibnamefont {LeBrun}}, \bibinfo {author} {\bibfnamefont {D.~C.}\ \bibnamefont {Moore}},\ and\ \bibinfo {author} {\bibfnamefont {J.~M.}\ \bibnamefont {Taylor}},\ }\bibfield  {title} {\bibinfo {title} {Collision-resolved pressure sensing},\ }\href {https://doi.org/10.1103/PhysRevA.109.042616} {\bibfield  {journal} {\bibinfo  {journal} {Physical Review A}\ }\textbf {\bibinfo {volume} {109}},\ \bibinfo {pages} {042616} (\bibinfo {year} {2024})}\BibitemShut {NoStop}%
\bibitem [{\citenamefont {Scherschligt}\ \emph {et~al.}(2018)\citenamefont {Scherschligt}, \citenamefont {Fedchak}, \citenamefont {Ahmed}, \citenamefont {Barker}, \citenamefont {Douglass}, \citenamefont {Eckel}, \citenamefont {Hanson}, \citenamefont {Hendricks}, \citenamefont {Klimov}, \citenamefont {Purdy}, \citenamefont {Ricker}, \citenamefont {Singh},\ and\ \citenamefont {Stone}}]{Scherschligt2018}%
  \BibitemOpen
  \bibfield  {author} {\bibinfo {author} {\bibfnamefont {J.}~\bibnamefont {Scherschligt}}, \bibinfo {author} {\bibfnamefont {J.~A.}\ \bibnamefont {Fedchak}}, \bibinfo {author} {\bibfnamefont {Z.}~\bibnamefont {Ahmed}}, \bibinfo {author} {\bibfnamefont {D.~S.}\ \bibnamefont {Barker}}, \bibinfo {author} {\bibfnamefont {K.}~\bibnamefont {Douglass}}, \bibinfo {author} {\bibfnamefont {S.}~\bibnamefont {Eckel}}, \bibinfo {author} {\bibfnamefont {E.}~\bibnamefont {Hanson}}, \bibinfo {author} {\bibfnamefont {J.}~\bibnamefont {Hendricks}}, \bibinfo {author} {\bibfnamefont {N.}~\bibnamefont {Klimov}}, \bibinfo {author} {\bibfnamefont {T.}~\bibnamefont {Purdy}}, \bibinfo {author} {\bibfnamefont {J.}~\bibnamefont {Ricker}}, \bibinfo {author} {\bibfnamefont {R.}~\bibnamefont {Singh}},\ and\ \bibinfo {author} {\bibfnamefont {J.}~\bibnamefont {Stone}},\ }\bibfield  {title} {\bibinfo {title} {Review {{Article}}: {{Quantum-based}} vacuum metrology at the {{National Institute}} of {{Standards}} and {{Technology}}},\ }\href
  {https://doi.org/10.1116/1.5033568} {\bibfield  {journal} {\bibinfo  {journal} {Journal of Vacuum Science \& Technology A}\ }\textbf {\bibinfo {volume} {36}},\ \bibinfo {pages} {040801} (\bibinfo {year} {2018})}\BibitemShut {NoStop}%
\bibitem [{\citenamefont {Blakemore}\ \emph {et~al.}(2020)\citenamefont {Blakemore}, \citenamefont {Martin}, \citenamefont {Fieguth}, \citenamefont {Kawasaki}, \citenamefont {Priel}, \citenamefont {Rider},\ and\ \citenamefont {Gratta}}]{Blakemore2020}%
  \BibitemOpen
  \bibfield  {author} {\bibinfo {author} {\bibfnamefont {C.~P.}\ \bibnamefont {Blakemore}}, \bibinfo {author} {\bibfnamefont {D.}~\bibnamefont {Martin}}, \bibinfo {author} {\bibfnamefont {A.}~\bibnamefont {Fieguth}}, \bibinfo {author} {\bibfnamefont {A.}~\bibnamefont {Kawasaki}}, \bibinfo {author} {\bibfnamefont {N.}~\bibnamefont {Priel}}, \bibinfo {author} {\bibfnamefont {A.~D.}\ \bibnamefont {Rider}},\ and\ \bibinfo {author} {\bibfnamefont {G.}~\bibnamefont {Gratta}},\ }\bibfield  {title} {\bibinfo {title} {Absolute pressure and gas species identification with an optically levitated rotor},\ }\href {https://doi.org/10.1116/1.5139638} {\bibfield  {journal} {\bibinfo  {journal} {Journal of Vacuum Science \& Technology B}\ }\textbf {\bibinfo {volume} {38}},\ \bibinfo {pages} {024201} (\bibinfo {year} {2020})}\BibitemShut {NoStop}%
\bibitem [{\citenamefont {Reinhardt}\ \emph {et~al.}(2024)\citenamefont {Reinhardt}, \citenamefont {Masalehdan}, \citenamefont {Croatto}, \citenamefont {Franke}, \citenamefont {Kunze}, \citenamefont {Schaffran}, \citenamefont {S{\"u}ltmann}, \citenamefont {Lindner},\ and\ \citenamefont {Schnabel}}]{reinhardt2024}%
  \BibitemOpen
  \bibfield  {author} {\bibinfo {author} {\bibfnamefont {C.}~\bibnamefont {Reinhardt}}, \bibinfo {author} {\bibfnamefont {H.}~\bibnamefont {Masalehdan}}, \bibinfo {author} {\bibfnamefont {S.}~\bibnamefont {Croatto}}, \bibinfo {author} {\bibfnamefont {A.}~\bibnamefont {Franke}}, \bibinfo {author} {\bibfnamefont {M.~B.~K.}\ \bibnamefont {Kunze}}, \bibinfo {author} {\bibfnamefont {J.}~\bibnamefont {Schaffran}}, \bibinfo {author} {\bibfnamefont {N.}~\bibnamefont {S{\"u}ltmann}}, \bibinfo {author} {\bibfnamefont {A.}~\bibnamefont {Lindner}},\ and\ \bibinfo {author} {\bibfnamefont {R.}~\bibnamefont {Schnabel}},\ }\bibfield  {title} {\bibinfo {title} {Self-{{Calibrating Gas Pressure Sensor}} with a 10-{{Decade Measurement Range}}},\ }\href {https://doi.org/10.1021/acsphotonics.3c01488} {\bibfield  {journal} {\bibinfo  {journal} {ACS Photonics}\ }\textbf {\bibinfo {volume} {11}},\ \bibinfo {pages} {1438} (\bibinfo {year} {2024})}\BibitemShut {NoStop}%
\bibitem [{\citenamefont {BIPM}\ \emph {et~al.}({\natexlab{a}})\citenamefont {BIPM}, \citenamefont {IEC}, \citenamefont {IFCC}, \citenamefont {ILAC}, \citenamefont {ISO}, \citenamefont {IUPAC}, \citenamefont {IUPAP},\ and\ \citenamefont {OIML}}]{JCGMVIM3}%
  \BibitemOpen
  \bibfield  {author} {\bibinfo {author} {\bibnamefont {BIPM}}, \bibinfo {author} {\bibnamefont {IEC}}, \bibinfo {author} {\bibnamefont {IFCC}}, \bibinfo {author} {\bibnamefont {ILAC}}, \bibinfo {author} {\bibnamefont {ISO}}, \bibinfo {author} {\bibnamefont {IUPAC}}, \bibinfo {author} {\bibnamefont {IUPAP}},\ and\ \bibinfo {author} {\bibnamefont {OIML}},\ }\href {https://www.bipm.org/documents/20126/2071204/JCGM_200_2012.pdf/f0e1ad45-d337-bbeb-53a6-15fe649d0ff1} {\bibinfo {title} {International vocabulary of metrology --- {B}asic and general concepts and associated terms ({VIM})}},\ \bibinfo {howpublished} {Joint Committee for Guides in Metrology, JCGM 200:2012. (3rd edition)} ({\natexlab{a}})\BibitemShut {NoStop}%
\bibitem [{\citenamefont {Comsa}\ \emph {et~al.}(1980)\citenamefont {Comsa}, \citenamefont {Fremerey}, \citenamefont {Lindenau}, \citenamefont {Messer},\ and\ \citenamefont {Roehl}}]{Comsa1980}%
  \BibitemOpen
  \bibfield  {author} {\bibinfo {author} {\bibfnamefont {G.}~\bibnamefont {Comsa}}, \bibinfo {author} {\bibfnamefont {J.~K.}\ \bibnamefont {Fremerey}}, \bibinfo {author} {\bibfnamefont {B.}~\bibnamefont {Lindenau}}, \bibinfo {author} {\bibfnamefont {G.}~\bibnamefont {Messer}},\ and\ \bibinfo {author} {\bibfnamefont {P.}~\bibnamefont {Roehl}},\ }\bibfield  {title} {\bibinfo {title} {Calibration of a spinning rotor gas friction gauge against a fundamental vacuum pressure standard},\ }\href {https://doi.org/10.1116/1.570531} {\bibfield  {journal} {\bibinfo  {journal} {Journal of Vacuum Science \& Technology}\ }\textbf {\bibinfo {volume} {17}},\ \bibinfo {pages} {642} (\bibinfo {year} {1980})}\BibitemShut {NoStop}%
\bibitem [{\citenamefont {Jousten}\ \emph {et~al.}(2021)\citenamefont {Jousten}, \citenamefont {Bernien}, \citenamefont {Boineau}, \citenamefont {Bundaleski}, \citenamefont {Illgen}, \citenamefont {Jenninger}, \citenamefont {J{\"o}nsson}, \citenamefont {{\v S}etina}, \citenamefont {Teodoro},\ and\ \citenamefont {Vi{\v c}ar}}]{jousten2021}%
  \BibitemOpen
  \bibfield  {author} {\bibinfo {author} {\bibfnamefont {K.}~\bibnamefont {Jousten}}, \bibinfo {author} {\bibfnamefont {M.}~\bibnamefont {Bernien}}, \bibinfo {author} {\bibfnamefont {F.}~\bibnamefont {Boineau}}, \bibinfo {author} {\bibfnamefont {N.}~\bibnamefont {Bundaleski}}, \bibinfo {author} {\bibfnamefont {C.}~\bibnamefont {Illgen}}, \bibinfo {author} {\bibfnamefont {B.}~\bibnamefont {Jenninger}}, \bibinfo {author} {\bibfnamefont {G.}~\bibnamefont {J{\"o}nsson}}, \bibinfo {author} {\bibfnamefont {J.}~\bibnamefont {{\v S}etina}}, \bibinfo {author} {\bibfnamefont {O.~M.}\ \bibnamefont {Teodoro}},\ and\ \bibinfo {author} {\bibfnamefont {M.}~\bibnamefont {Vi{\v c}ar}},\ }\bibfield  {title} {\bibinfo {title} {Electrons on a straight path: {{A}} novel ionisation vacuum gauge suitable as reference standard},\ }\href {https://doi.org/10.1016/j.vacuum.2021.110239} {\bibfield  {journal} {\bibinfo  {journal} {Vacuum}\ }\textbf {\bibinfo {volume} {189}},\ \bibinfo {pages} {110239} (\bibinfo {year}
  {2021})}\BibitemShut {NoStop}%
\bibitem [{\citenamefont {Egan}\ \emph {et~al.}(2015)\citenamefont {Egan}, \citenamefont {Stone}, \citenamefont {Hendricks}, \citenamefont {Ricker}, \citenamefont {Scace},\ and\ \citenamefont {Strouse}}]{egan2015}%
  \BibitemOpen
  \bibfield  {author} {\bibinfo {author} {\bibfnamefont {P.~F.}\ \bibnamefont {Egan}}, \bibinfo {author} {\bibfnamefont {J.~A.}\ \bibnamefont {Stone}}, \bibinfo {author} {\bibfnamefont {J.~H.}\ \bibnamefont {Hendricks}}, \bibinfo {author} {\bibfnamefont {J.~E.}\ \bibnamefont {Ricker}}, \bibinfo {author} {\bibfnamefont {G.~E.}\ \bibnamefont {Scace}},\ and\ \bibinfo {author} {\bibfnamefont {G.~F.}\ \bibnamefont {Strouse}},\ }\bibfield  {title} {\bibinfo {title} {Performance of a dual {{Fabry}} -- {{Perot}} cavity refractometer},\ }\href@noop {} {\bibfield  {journal} {\bibinfo  {journal} {Optics Letters}\ }\textbf {\bibinfo {volume} {40}},\ \bibinfo {pages} {3945} (\bibinfo {year} {2015})}\BibitemShut {NoStop}%
\bibitem [{\citenamefont {Egan}\ \emph {et~al.}(2016)\citenamefont {Egan}, \citenamefont {Stone}, \citenamefont {Ricker},\ and\ \citenamefont {Hendricks}}]{egan2016}%
  \BibitemOpen
  \bibfield  {author} {\bibinfo {author} {\bibfnamefont {P.~F.}\ \bibnamefont {Egan}}, \bibinfo {author} {\bibfnamefont {J.~A.}\ \bibnamefont {Stone}}, \bibinfo {author} {\bibfnamefont {J.~E.}\ \bibnamefont {Ricker}},\ and\ \bibinfo {author} {\bibfnamefont {J.~H.}\ \bibnamefont {Hendricks}},\ }\bibfield  {title} {\bibinfo {title} {Comparison measurements of low-pressure between a laser refractometer and ultrasonic manometer},\ }\href {https://doi.org/10.1063/1.4949504} {\bibfield  {journal} {\bibinfo  {journal} {Review of Scientific Instruments}\ }\textbf {\bibinfo {volume} {87}},\ \bibinfo {pages} {053113} (\bibinfo {year} {2016})}\BibitemShut {NoStop}%
\bibitem [{\citenamefont {Barker}\ \emph {et~al.}(2023)\citenamefont {Barker}, \citenamefont {Fedchak}, \citenamefont {K{\l}os}, \citenamefont {Scherschligt}, \citenamefont {Sheikh}, \citenamefont {Tiesinga},\ and\ \citenamefont {Eckel}}]{barker2023}%
  \BibitemOpen
  \bibfield  {author} {\bibinfo {author} {\bibfnamefont {D.~S.}\ \bibnamefont {Barker}}, \bibinfo {author} {\bibfnamefont {J.~A.}\ \bibnamefont {Fedchak}}, \bibinfo {author} {\bibfnamefont {J.}~\bibnamefont {K{\l}os}}, \bibinfo {author} {\bibfnamefont {J.}~\bibnamefont {Scherschligt}}, \bibinfo {author} {\bibfnamefont {A.~A.}\ \bibnamefont {Sheikh}}, \bibinfo {author} {\bibfnamefont {E.}~\bibnamefont {Tiesinga}},\ and\ \bibinfo {author} {\bibfnamefont {S.~P.}\ \bibnamefont {Eckel}},\ }\bibfield  {title} {\bibinfo {title} {Accurate measurement of the loss rate of cold atoms due to background gas collisions for the quantum-based cold atom vacuum standard},\ }\href {https://doi.org/10.1116/5.0147686} {\bibfield  {journal} {\bibinfo  {journal} {AVS Quantum Science}\ }\textbf {\bibinfo {volume} {5}},\ \bibinfo {pages} {035001} (\bibinfo {year} {2023})}\BibitemShut {NoStop}%
\bibitem [{\citenamefont {Cavalleri}\ \emph {et~al.}(2010)\citenamefont {Cavalleri}, \citenamefont {Ciani}, \citenamefont {Dolesi}, \citenamefont {Hueller}, \citenamefont {Nicolodi}, \citenamefont {Tombolato}, \citenamefont {Vitale}, \citenamefont {Wass},\ and\ \citenamefont {Weber}}]{Cavalleri2010}%
  \BibitemOpen
  \bibfield  {author} {\bibinfo {author} {\bibfnamefont {A.}~\bibnamefont {Cavalleri}}, \bibinfo {author} {\bibfnamefont {G.}~\bibnamefont {Ciani}}, \bibinfo {author} {\bibfnamefont {R.}~\bibnamefont {Dolesi}}, \bibinfo {author} {\bibfnamefont {M.}~\bibnamefont {Hueller}}, \bibinfo {author} {\bibfnamefont {D.}~\bibnamefont {Nicolodi}}, \bibinfo {author} {\bibfnamefont {D.}~\bibnamefont {Tombolato}}, \bibinfo {author} {\bibfnamefont {S.}~\bibnamefont {Vitale}}, \bibinfo {author} {\bibfnamefont {P.~J.}\ \bibnamefont {Wass}},\ and\ \bibinfo {author} {\bibfnamefont {W.~J.}\ \bibnamefont {Weber}},\ }\bibfield  {title} {\bibinfo {title} {Gas damping force noise on a macroscopic test body in an infinite gas reservoir},\ }\href {https://doi.org/10.1016/j.physleta.2010.06.041} {\bibfield  {journal} {\bibinfo  {journal} {Physics Letters A}\ }\textbf {\bibinfo {volume} {374}},\ \bibinfo {pages} {3365} (\bibinfo {year} {2010})}\BibitemShut {NoStop}%
\bibitem [{\citenamefont {Martinetz}\ \emph {et~al.}(2018)\citenamefont {Martinetz}, \citenamefont {Hornberger},\ and\ \citenamefont {Stickler}}]{Martinetz2018}%
  \BibitemOpen
  \bibfield  {author} {\bibinfo {author} {\bibfnamefont {L.}~\bibnamefont {Martinetz}}, \bibinfo {author} {\bibfnamefont {K.}~\bibnamefont {Hornberger}},\ and\ \bibinfo {author} {\bibfnamefont {B.~A.}\ \bibnamefont {Stickler}},\ }\bibfield  {title} {\bibinfo {title} {Gas-induced friction and diffusion of rigid rotors},\ }\href {https://doi.org/10.1103/PhysRevE.97.052112} {\bibfield  {journal} {\bibinfo  {journal} {Physical Review E}\ }\textbf {\bibinfo {volume} {97}},\ \bibinfo {pages} {052112} (\bibinfo {year} {2018})}\BibitemShut {NoStop}%
\bibitem [{\citenamefont {Reinhardt}\ \emph {et~al.}(2016)\citenamefont {Reinhardt}, \citenamefont {M{\"u}ller}, \citenamefont {Bourassa},\ and\ \citenamefont {Sankey}}]{Reinhardt2016}%
  \BibitemOpen
  \bibfield  {author} {\bibinfo {author} {\bibfnamefont {C.}~\bibnamefont {Reinhardt}}, \bibinfo {author} {\bibfnamefont {T.}~\bibnamefont {M{\"u}ller}}, \bibinfo {author} {\bibfnamefont {A.}~\bibnamefont {Bourassa}},\ and\ \bibinfo {author} {\bibfnamefont {J.~C.}\ \bibnamefont {Sankey}},\ }\bibfield  {title} {\bibinfo {title} {Ultralow-{{Noise SiN Trampoline Resonators}} for {{Sensing}} and {{Optomechanics}}},\ }\href {https://doi.org/10.1103/PhysRevX.6.021001} {\bibfield  {journal} {\bibinfo  {journal} {Physical Review X}\ }\textbf {\bibinfo {volume} {6}},\ \bibinfo {pages} {021001} (\bibinfo {year} {2016})}\BibitemShut {NoStop}%
\bibitem [{\citenamefont {Norte}\ \emph {et~al.}(2016)\citenamefont {Norte}, \citenamefont {Moura},\ and\ \citenamefont {Gr{\"o}blacher}}]{Norte2016}%
  \BibitemOpen
  \bibfield  {author} {\bibinfo {author} {\bibfnamefont {R.~A.}\ \bibnamefont {Norte}}, \bibinfo {author} {\bibfnamefont {J.~P.}\ \bibnamefont {Moura}},\ and\ \bibinfo {author} {\bibfnamefont {S.}~\bibnamefont {Gr{\"o}blacher}},\ }\bibfield  {title} {\bibinfo {title} {Mechanical {{Resonators}} for {{Quantum Optomechanics Experiments}} at {{Room Temperature}}},\ }\href {https://doi.org/10.1103/PhysRevLett.116.147202} {\bibfield  {journal} {\bibinfo  {journal} {Physical Review Letters}\ }\textbf {\bibinfo {volume} {116}},\ \bibinfo {pages} {147202} (\bibinfo {year} {2016})}\BibitemShut {NoStop}%
\bibitem [{dis()}]{disclaimer}%
  \BibitemOpen
  \href@noop {} {}\bibinfo {note} {Certain commercial equipment, instruments, or materials are identified for reference purposes only. Such identification is not intended to imply recommendation or endorsement by NIST, nor is it intended to imply that the materials or equipment identified are necessarily the best available for the purpose.}\BibitemShut {Stop}%
\bibitem [{\citenamefont {BIPM}\ \emph {et~al.}({\natexlab{b}})\citenamefont {BIPM}, \citenamefont {IEC}, \citenamefont {IFCC}, \citenamefont {ILAC}, \citenamefont {ISO}, \citenamefont {IUPAC}, \citenamefont {IUPAP},\ and\ \citenamefont {OIML}}]{JCGMGUM}%
  \BibitemOpen
  \bibfield  {author} {\bibinfo {author} {\bibnamefont {BIPM}}, \bibinfo {author} {\bibnamefont {IEC}}, \bibinfo {author} {\bibnamefont {IFCC}}, \bibinfo {author} {\bibnamefont {ILAC}}, \bibinfo {author} {\bibnamefont {ISO}}, \bibinfo {author} {\bibnamefont {IUPAC}}, \bibinfo {author} {\bibnamefont {IUPAP}},\ and\ \bibinfo {author} {\bibnamefont {OIML}},\ }\href {https://www.bipm.org/documents/20126/2071204/JCGM\_100\_2008\_E.pdf/cb0ef43f-baa5-11cf-3f85-4dcd86f77bd6} {\bibinfo {title} {Evaluation of measurement data --- {G}uide to the expression of uncertainty in measurement}},\ \bibinfo {howpublished} {Joint Committee for Guides in Metrology, JCGM 100:2008} ({\natexlab{b}})\BibitemShut {NoStop}%
\bibitem [{\citenamefont {Miller}(1997)}]{miller1997}%
  \BibitemOpen
  \bibfield  {author} {\bibinfo {author} {\bibfnamefont {A.~P.}\ \bibnamefont {Miller}},\ }\bibfield  {title} {\bibinfo {title} {Measurement performance of capacitance diaphragm gages and alternative low-pressure transducers},\ }\href@noop {} {\bibfield  {journal} {\bibinfo  {journal} {NCSL Workshop \& Symposium}\ ,\ \bibinfo {pages} {287}} (\bibinfo {year} {1997})}\BibitemShut {NoStop}%
\bibitem [{\citenamefont {Ricker}\ \emph {et~al.}(2017)\citenamefont {Ricker}, \citenamefont {Hendricks}, \citenamefont {Bock}, \citenamefont {Dominik}, \citenamefont {Kobata}, \citenamefont {Torres},\ and\ \citenamefont {Sadkovskaya}}]{ricker2017}%
  \BibitemOpen
  \bibfield  {author} {\bibinfo {author} {\bibfnamefont {J.}~\bibnamefont {Ricker}}, \bibinfo {author} {\bibfnamefont {J.}~\bibnamefont {Hendricks}}, \bibinfo {author} {\bibfnamefont {T.}~\bibnamefont {Bock}}, \bibinfo {author} {\bibfnamefont {k.}~\bibnamefont {Dominik}}, \bibinfo {author} {\bibfnamefont {T.}~\bibnamefont {Kobata}}, \bibinfo {author} {\bibfnamefont {J.}~\bibnamefont {Torres}},\ and\ \bibinfo {author} {\bibfnamefont {I.}~\bibnamefont {Sadkovskaya}},\ }\bibfield  {title} {\bibinfo {title} {Final report on the key comparison {{CCM}}.{{P-K4}}.2012 in absolute pressure from 1 {{Pa}} to 10 {{kPa}}},\ }\href {https://doi.org/10.1088/0026-1394/54/1A/07002} {\bibfield  {journal} {\bibinfo  {journal} {Metrologia}\ }\textbf {\bibinfo {volume} {54}},\ \bibinfo {pages} {07002} (\bibinfo {year} {2017})}\BibitemShut {NoStop}%
\bibitem [{\citenamefont {Chakram}\ \emph {et~al.}(2014)\citenamefont {Chakram}, \citenamefont {Patil}, \citenamefont {Chang},\ and\ \citenamefont {Vengalattore}}]{Chakram2014}%
  \BibitemOpen
  \bibfield  {author} {\bibinfo {author} {\bibfnamefont {S.}~\bibnamefont {Chakram}}, \bibinfo {author} {\bibfnamefont {Y.~S.}\ \bibnamefont {Patil}}, \bibinfo {author} {\bibfnamefont {L.}~\bibnamefont {Chang}},\ and\ \bibinfo {author} {\bibfnamefont {M.}~\bibnamefont {Vengalattore}},\ }\bibfield  {title} {\bibinfo {title} {Dissipation in ultrahigh quality factor sin membrane resonators},\ }\href {https://doi.org/10.1103/physrevlett.112.127201} {\bibfield  {journal} {\bibinfo  {journal} {Physical Review Letters}\ }\textbf {\bibinfo {volume} {112}},\ \bibinfo {pages} {127201} (\bibinfo {year} {2014})}\BibitemShut {NoStop}%
\bibitem [{\citenamefont {Tsaturyan}\ \emph {et~al.}(2017)\citenamefont {Tsaturyan}, \citenamefont {Barg}, \citenamefont {Polzik},\ and\ \citenamefont {Schliesser}}]{Tsaturyan2017}%
  \BibitemOpen
  \bibfield  {author} {\bibinfo {author} {\bibfnamefont {Y.}~\bibnamefont {Tsaturyan}}, \bibinfo {author} {\bibfnamefont {A.}~\bibnamefont {Barg}}, \bibinfo {author} {\bibfnamefont {E.~S.}\ \bibnamefont {Polzik}},\ and\ \bibinfo {author} {\bibfnamefont {A.}~\bibnamefont {Schliesser}},\ }\bibfield  {title} {\bibinfo {title} {Ultracoherent nanomechanical resonators via soft clamping and dissipation dilution},\ }\href {https://doi.org/10.1038/nnano.2017.101} {\bibfield  {journal} {\bibinfo  {journal} {Nature Nanotechnology}\ }\textbf {\bibinfo {volume} {12}},\ \bibinfo {pages} {776} (\bibinfo {year} {2017})}\BibitemShut {NoStop}%
\bibitem [{\citenamefont {Christian}(1966)}]{Christian1966}%
  \BibitemOpen
  \bibfield  {author} {\bibinfo {author} {\bibfnamefont {R.~G.}\ \bibnamefont {Christian}},\ }\bibfield  {title} {\bibinfo {title} {The theory of oscillating-vane vacuum gauges},\ }\href {https://doi.org/10.1016/0042-207X(66)91162-6} {\bibfield  {journal} {\bibinfo  {journal} {Vacuum}\ }\textbf {\bibinfo {volume} {16}},\ \bibinfo {pages} {175} (\bibinfo {year} {1966})}\BibitemShut {NoStop}%
\bibitem [{\citenamefont {Newell}(1968)}]{newell1968}%
  \BibitemOpen
  \bibfield  {author} {\bibinfo {author} {\bibfnamefont {W.~E.}\ \bibnamefont {Newell}},\ }\bibfield  {title} {\bibinfo {title} {Miniaturization of {{Tuning Forks}}},\ }\href {https://doi.org/10.1126/science.161.3848.1320} {\bibfield  {journal} {\bibinfo  {journal} {Science}\ }\textbf {\bibinfo {volume} {161}},\ \bibinfo {pages} {1320} (\bibinfo {year} {1968})}\BibitemShut {NoStop}%
\bibitem [{\citenamefont {Hosaka}\ \emph {et~al.}(1995)\citenamefont {Hosaka}, \citenamefont {Itao},\ and\ \citenamefont {Kuroda}}]{hosaka1995}%
  \BibitemOpen
  \bibfield  {author} {\bibinfo {author} {\bibfnamefont {H.}~\bibnamefont {Hosaka}}, \bibinfo {author} {\bibfnamefont {K.}~\bibnamefont {Itao}},\ and\ \bibinfo {author} {\bibfnamefont {S.}~\bibnamefont {Kuroda}},\ }\bibfield  {title} {\bibinfo {title} {Damping characteristics of beam-shaped micro-oscillators},\ }\href {https://doi.org/10.1016/0924-4247(95)01003-J} {\bibfield  {journal} {\bibinfo  {journal} {Sensors and Actuators A}\ }\textbf {\bibinfo {volume} {49}},\ \bibinfo {pages} {87} (\bibinfo {year} {1995})}\BibitemShut {NoStop}%
\bibitem [{\citenamefont {L{\"u}bbe}\ \emph {et~al.}(2011)\citenamefont {L{\"u}bbe}, \citenamefont {Temmen}, \citenamefont {Schnieder},\ and\ \citenamefont {Reichling}}]{lubbe2011}%
  \BibitemOpen
  \bibfield  {author} {\bibinfo {author} {\bibfnamefont {J.}~\bibnamefont {L{\"u}bbe}}, \bibinfo {author} {\bibfnamefont {M.}~\bibnamefont {Temmen}}, \bibinfo {author} {\bibfnamefont {H.}~\bibnamefont {Schnieder}},\ and\ \bibinfo {author} {\bibfnamefont {M.}~\bibnamefont {Reichling}},\ }\bibfield  {title} {\bibinfo {title} {Measurement and modelling of non-contact atomic force microscope cantilever properties from ultra-high vacuum to normal pressure conditions},\ }\href {https://doi.org/10.1088/0957-0233/22/5/055501} {\bibfield  {journal} {\bibinfo  {journal} {Measurement Science and Technology}\ }\textbf {\bibinfo {volume} {22}},\ \bibinfo {pages} {055501} (\bibinfo {year} {2011})}\BibitemShut {NoStop}%
\bibitem [{Note1()}]{Note1}%
  \BibitemOpen
  \bibinfo {note} {Eventually, the size of the chamber enclosing the oscillator will also affect the damping rate, see Ref.~\cite {reich1982, Fremerey1985}.}\BibitemShut {Stop}%
\bibitem [{\citenamefont {Ramsay}(1956)}]{RamsayMolBeams}%
  \BibitemOpen
  \bibfield  {author} {\bibinfo {author} {\bibfnamefont {N.~F.}\ \bibnamefont {Ramsay}},\ }\href@noop {} {\emph {\bibinfo {title} {Molecular Beams}}}\ (\bibinfo  {publisher} {Oxford University Press Inc.},\ \bibinfo {address} {New York},\ \bibinfo {year} {1956})\BibitemShut {NoStop}%
\bibitem [{\citenamefont {Markwitz}\ \emph {et~al.}(1993)\citenamefont {Markwitz}, \citenamefont {Baumann}, \citenamefont {Krimmel}, \citenamefont {Bethge},\ and\ \citenamefont {Misaelides}}]{markwitz1993}%
  \BibitemOpen
  \bibfield  {author} {\bibinfo {author} {\bibfnamefont {A.}~\bibnamefont {Markwitz}}, \bibinfo {author} {\bibfnamefont {H.}~\bibnamefont {Baumann}}, \bibinfo {author} {\bibfnamefont {E.~F.}\ \bibnamefont {Krimmel}}, \bibinfo {author} {\bibfnamefont {K.}~\bibnamefont {Bethge}},\ and\ \bibinfo {author} {\bibfnamefont {P.}~\bibnamefont {Misaelides}},\ }\bibfield  {title} {\bibinfo {title} {Characterisation of thin sputtered silicon nitride films by {{NRA}}, {{ERDA}}, {{RBS}} and {{SEM}}},\ }\href {https://doi.org/10.1007/BF00321408} {\bibfield  {journal} {\bibinfo  {journal} {Fresenius' Journal of Analytical Chemistry}\ }\textbf {\bibinfo {volume} {346}},\ \bibinfo {pages} {177} (\bibinfo {year} {1993})}\BibitemShut {NoStop}%
\bibitem [{\citenamefont {Huszank}\ \emph {et~al.}(2016)\citenamefont {Huszank}, \citenamefont {Csedreki}, \citenamefont {Kert{\'e}sz},\ and\ \citenamefont {T{\"o}r{\"o}k}}]{huszank2016}%
  \BibitemOpen
  \bibfield  {author} {\bibinfo {author} {\bibfnamefont {R.}~\bibnamefont {Huszank}}, \bibinfo {author} {\bibfnamefont {L.}~\bibnamefont {Csedreki}}, \bibinfo {author} {\bibfnamefont {Z.}~\bibnamefont {Kert{\'e}sz}},\ and\ \bibinfo {author} {\bibfnamefont {Z.}~\bibnamefont {T{\"o}r{\"o}k}},\ }\bibfield  {title} {\bibinfo {title} {Determination of the density of silicon--nitride thin films by ion-beam analytical techniques ({{RBS}}, {{PIXE}}, {{STIM}})},\ }\href {https://doi.org/10.1007/s10967-015-4102-9} {\bibfield  {journal} {\bibinfo  {journal} {Journal of Radioanalytical and Nuclear Chemistry}\ }\textbf {\bibinfo {volume} {307}},\ \bibinfo {pages} {341} (\bibinfo {year} {2016})}\BibitemShut {NoStop}%
\bibitem [{\citenamefont {Trott}\ \emph {et~al.}(2011)\citenamefont {Trott}, \citenamefont {Casta{\~n}eda}, \citenamefont {Torczynski}, \citenamefont {Gallis},\ and\ \citenamefont {Rader}}]{trott2011}%
  \BibitemOpen
  \bibfield  {author} {\bibinfo {author} {\bibfnamefont {W.~M.}\ \bibnamefont {Trott}}, \bibinfo {author} {\bibfnamefont {J.~N.}\ \bibnamefont {Casta{\~n}eda}}, \bibinfo {author} {\bibfnamefont {J.~R.}\ \bibnamefont {Torczynski}}, \bibinfo {author} {\bibfnamefont {M.~A.}\ \bibnamefont {Gallis}},\ and\ \bibinfo {author} {\bibfnamefont {D.~J.}\ \bibnamefont {Rader}},\ }\bibfield  {title} {\bibinfo {title} {An experimental assembly for precise measurement of thermal accommodation coefficients},\ }\href {https://doi.org/10.1063/1.3571269} {\bibfield  {journal} {\bibinfo  {journal} {Review of Scientific Instruments}\ }\textbf {\bibinfo {volume} {82}},\ \bibinfo {pages} {035120} (\bibinfo {year} {2011})}\BibitemShut {NoStop}%
\bibitem [{\citenamefont {Sharipov}\ and\ \citenamefont {Moldover}(2016)}]{sharipov2016}%
  \BibitemOpen
  \bibfield  {author} {\bibinfo {author} {\bibfnamefont {F.}~\bibnamefont {Sharipov}}\ and\ \bibinfo {author} {\bibfnamefont {M.~R.}\ \bibnamefont {Moldover}},\ }\bibfield  {title} {\bibinfo {title} {Energy accommodation coefficient extracted from acoustic resonator experiments},\ }\href {https://doi.org/10.1116/1.4966620} {\bibfield  {journal} {\bibinfo  {journal} {Journal of Vacuum Science \& Technology A}\ }\textbf {\bibinfo {volume} {34}},\ \bibinfo {pages} {061604} (\bibinfo {year} {2016})}\BibitemShut {NoStop}%
\bibitem [{Note2()}]{Note2}%
  \BibitemOpen
  \bibinfo {note} {When SiN~3 is exposed to uncontrolled environmental changes (\protect \textit {e.g.} long-term exposure to atmosphere), the drift in \(\Gamma _0\) is no longer consistent with zero, but it remains equivalent to a pressure drift more than \(10\) times lower than the secondary standard}\BibitemShut {NoStop}%
\bibitem [{\citenamefont {Xu}\ \emph {et~al.}(2023)\citenamefont {Xu}, \citenamefont {Shin}, \citenamefont {Sberna}, \citenamefont {van~der Kolk}, \citenamefont {Cupertino}, \citenamefont {Bessa},\ and\ \citenamefont {Norte}}]{Xu2023}%
  \BibitemOpen
  \bibfield  {author} {\bibinfo {author} {\bibfnamefont {M.}~\bibnamefont {Xu}}, \bibinfo {author} {\bibfnamefont {D.}~\bibnamefont {Shin}}, \bibinfo {author} {\bibfnamefont {P.~M.}\ \bibnamefont {Sberna}}, \bibinfo {author} {\bibfnamefont {R.}~\bibnamefont {van~der Kolk}}, \bibinfo {author} {\bibfnamefont {A.}~\bibnamefont {Cupertino}}, \bibinfo {author} {\bibfnamefont {M.~A.}\ \bibnamefont {Bessa}},\ and\ \bibinfo {author} {\bibfnamefont {R.~A.}\ \bibnamefont {Norte}},\ }\bibfield  {title} {\bibinfo {title} {High‐strength amorphous silicon carbide for nanomechanics},\ }\href {https://doi.org/10.1002/adma.202306513} {\bibfield  {journal} {\bibinfo  {journal} {Advanced Materials}\ }\textbf {\bibinfo {volume} {36}},\ \bibinfo {pages} {2306513} (\bibinfo {year} {2023})}\BibitemShut {NoStop}%
\bibitem [{zot()}]{zotero-2534}%
  \BibitemOpen
  \href@noop {} {\bibinfo {title} {X-ray {{Microscopy Windows Specification Sheet}}}},\ \bibinfo {howpublished} {https://www.norcada.com/wp-content/uploads/2013/10/Xray-Window-Specsheet.pdf}\BibitemShut {NoStop}%
\bibitem [{\citenamefont {Pickering}\ \emph {et~al.}(1990)\citenamefont {Pickering}, \citenamefont {Taylor}, \citenamefont {Keeley},\ and\ \citenamefont {Graves}}]{pickering1990}%
  \BibitemOpen
  \bibfield  {author} {\bibinfo {author} {\bibfnamefont {M.~A.}\ \bibnamefont {Pickering}}, \bibinfo {author} {\bibfnamefont {R.~L.}\ \bibnamefont {Taylor}}, \bibinfo {author} {\bibfnamefont {J.~T.}\ \bibnamefont {Keeley}},\ and\ \bibinfo {author} {\bibfnamefont {G.~A.}\ \bibnamefont {Graves}},\ }\bibfield  {title} {\bibinfo {title} {Chemically vapor deposited silicon carbide ({{SiC}}) for optical applications},\ }\href@noop {} {\bibfield  {journal} {\bibinfo  {journal} {Nuclear Instruments and Methods in Physics Research A}\ }\textbf {\bibinfo {volume} {291}},\ \bibinfo {pages} {95} (\bibinfo {year} {1990})}\BibitemShut {NoStop}%
\bibitem [{\citenamefont {Goela}\ \emph {et~al.}(1991)\citenamefont {Goela}, \citenamefont {Pickering}, \citenamefont {Taylor}, \citenamefont {Murray},\ and\ \citenamefont {Lompado}}]{goela1991}%
  \BibitemOpen
  \bibfield  {author} {\bibinfo {author} {\bibfnamefont {J.~S.}\ \bibnamefont {Goela}}, \bibinfo {author} {\bibfnamefont {M.~A.}\ \bibnamefont {Pickering}}, \bibinfo {author} {\bibfnamefont {R.~L.}\ \bibnamefont {Taylor}}, \bibinfo {author} {\bibfnamefont {B.~W.}\ \bibnamefont {Murray}},\ and\ \bibinfo {author} {\bibfnamefont {A.}~\bibnamefont {Lompado}},\ }\bibfield  {title} {\bibinfo {title} {Properties of chemical-vapor-deposited silicon carbide for optics applications in severe environments},\ }\href {https://doi.org/10.1364/AO.30.003166} {\bibfield  {journal} {\bibinfo  {journal} {Applied Optics}\ }\textbf {\bibinfo {volume} {30}},\ \bibinfo {pages} {3166} (\bibinfo {year} {1991})}\BibitemShut {NoStop}%
\bibitem [{\citenamefont {Hauer}\ \emph {et~al.}(2013)\citenamefont {Hauer}, \citenamefont {Doolin}, \citenamefont {Beach},\ and\ \citenamefont {Davis}}]{Hauer2013}%
  \BibitemOpen
  \bibfield  {author} {\bibinfo {author} {\bibfnamefont {B.}~\bibnamefont {Hauer}}, \bibinfo {author} {\bibfnamefont {C.}~\bibnamefont {Doolin}}, \bibinfo {author} {\bibfnamefont {K.}~\bibnamefont {Beach}},\ and\ \bibinfo {author} {\bibfnamefont {J.}~\bibnamefont {Davis}},\ }\bibfield  {title} {\bibinfo {title} {A general procedure for thermomechanical calibration of nano/micro-mechanical resonators},\ }\href {https://doi.org/10.1016/j.aop.2013.08.003} {\bibfield  {journal} {\bibinfo  {journal} {Annals of Physics}\ }\textbf {\bibinfo {volume} {339}},\ \bibinfo {pages} {181} (\bibinfo {year} {2013})}\BibitemShut {NoStop}%
\bibitem [{\citenamefont {Schmelling}(1995)}]{schmelling1995}%
  \BibitemOpen
  \bibfield  {author} {\bibinfo {author} {\bibfnamefont {M.}~\bibnamefont {Schmelling}},\ }\bibfield  {title} {\bibinfo {title} {Averaging correlated data},\ }\href {https://doi.org/10.1088/0031-8949/51/6/002} {\bibfield  {journal} {\bibinfo  {journal} {Physica Scripta}\ }\textbf {\bibinfo {volume} {51}},\ \bibinfo {pages} {676} (\bibinfo {year} {1995})}\BibitemShut {NoStop}%
\bibitem [{\citenamefont {Bevington}\ and\ \citenamefont {Robinson}(1992)}]{Bevington1992}%
  \BibitemOpen
  \bibfield  {author} {\bibinfo {author} {\bibfnamefont {P.}~\bibnamefont {Bevington}}\ and\ \bibinfo {author} {\bibfnamefont {D.}~\bibnamefont {Robinson}},\ }\href {https://books.google.com/books?id=oAovAQAAIAAJ} {\emph {\bibinfo {title} {Data Reduction and Error Analysis for the Physical Sciences}}},\ \bibinfo {series} {Book and Disk}\ No.\ \bibinfo {number} {v. 1}\ (\bibinfo  {publisher} {McGraw-Hill},\ \bibinfo {year} {1992})\BibitemShut {NoStop}%
\bibitem [{\citenamefont {{Foreman-Mackey}}\ \emph {et~al.}(2013)\citenamefont {{Foreman-Mackey}}, \citenamefont {Hogg}, \citenamefont {Lang},\ and\ \citenamefont {Goodman}}]{foreman-mackey2013}%
  \BibitemOpen
  \bibfield  {author} {\bibinfo {author} {\bibfnamefont {D.}~\bibnamefont {{Foreman-Mackey}}}, \bibinfo {author} {\bibfnamefont {D.~W.}\ \bibnamefont {Hogg}}, \bibinfo {author} {\bibfnamefont {D.}~\bibnamefont {Lang}},\ and\ \bibinfo {author} {\bibfnamefont {J.}~\bibnamefont {Goodman}},\ }\bibfield  {title} {\bibinfo {title} {Emcee: {{The MCMC Hammer}}},\ }\href {https://doi.org/10.1086/670067} {\bibfield  {journal} {\bibinfo  {journal} {Publications of the Astronomical Society of the Pacific}\ }\textbf {\bibinfo {volume} {125}},\ \bibinfo {pages} {306} (\bibinfo {year} {2013})}\BibitemShut {NoStop}%
\bibitem [{\citenamefont {Takaishi}\ and\ \citenamefont {Sensui}(1963)}]{takaishi1963}%
  \BibitemOpen
  \bibfield  {author} {\bibinfo {author} {\bibfnamefont {T.}~\bibnamefont {Takaishi}}\ and\ \bibinfo {author} {\bibfnamefont {Y.}~\bibnamefont {Sensui}},\ }\bibfield  {title} {\bibinfo {title} {Thermal transpiration effect of hydrogen, rare gases and methane},\ }\href {https://doi.org/10.1039/tf9635902503} {\bibfield  {journal} {\bibinfo  {journal} {Transactions of the Faraday Society}\ }\textbf {\bibinfo {volume} {59}},\ \bibinfo {pages} {2503} (\bibinfo {year} {1963})}\BibitemShut {NoStop}%
\bibitem [{\citenamefont {Poulter}\ \emph {et~al.}(1983)\citenamefont {Poulter}, \citenamefont {Rodgers}, \citenamefont {Nash}, \citenamefont {Thompson},\ and\ \citenamefont {Perkin}}]{poulter1983}%
  \BibitemOpen
  \bibfield  {author} {\bibinfo {author} {\bibfnamefont {K.~F.}\ \bibnamefont {Poulter}}, \bibinfo {author} {\bibfnamefont {M.-J.}\ \bibnamefont {Rodgers}}, \bibinfo {author} {\bibfnamefont {P.~J.}\ \bibnamefont {Nash}}, \bibinfo {author} {\bibfnamefont {T.~J.}\ \bibnamefont {Thompson}},\ and\ \bibinfo {author} {\bibfnamefont {M.~P.}\ \bibnamefont {Perkin}},\ }\bibfield  {title} {\bibinfo {title} {Thermal transpiration correction in capacitance manometers},\ }\href {https://doi.org/10.1016/0042-207X(83)90098-2} {\bibfield  {journal} {\bibinfo  {journal} {Vacuum}\ }\textbf {\bibinfo {volume} {33}},\ \bibinfo {pages} {311} (\bibinfo {year} {1983})}\BibitemShut {NoStop}%
\bibitem [{\citenamefont {Reich}(1982)}]{reich1982}%
  \BibitemOpen
  \bibfield  {author} {\bibinfo {author} {\bibfnamefont {G.}~\bibnamefont {Reich}},\ }\bibfield  {title} {\bibinfo {title} {Spinning rotor viscosity gauge: {{A}} transfer standard for the laboratory or an accurate gauge for vacuum process control},\ }\href {https://doi.org/10.1116/1.571592} {\bibfield  {journal} {\bibinfo  {journal} {Journal of Vacuum Science and Technology}\ }\textbf {\bibinfo {volume} {20}},\ \bibinfo {pages} {1148} (\bibinfo {year} {1982})}\BibitemShut {NoStop}%
\end{thebibliography}%

\end{document}